\documentclass{aa}  

\usepackage{graphicx}
\usepackage{txfonts}
\usepackage[pdfpagelabels=false]{hyperref}
\hypersetup{colorlinks=true,linkcolor=blue,citecolor=blue,filecolor=blue,urlcolor=blue}

\usepackage{amsmath}    
\usepackage{amssymb}    

\newcommand\rs[1]{_\mathrm{_{#1}}}
\newcommand\Vis{\gamma\rs{12}}
\newcommand\VisSq{|\gamma\rs{12}|^2}
\newcommand\SNgam{(S/N)\rs{|\Vis|^2}}
\newcommand\SNgamfir{(S/N)\rs{|\Vis|_{,1}^2}}
\newcommand\SNgamsec{(S/N)\rs{|\Vis|_{,2}^2}}

\newcommand\SNtheta{(S/N)\rs{\theta}}
\newcommand\etp{\eta\rs{1p}}
\newcommand\epsp{\epsilon\rs{1p}}
\newcommand\etpp{\eta\rs{2p}}
\newcommand\epspp{\epsilon\rs{2p}}
\newcommand\epsppp{\epsilon\rs{3p}}
\newcommand\etzbc{\eta\rs{ZBC}}
\newcommand\epszbc{\epsilon\rs{ZBC}}
\newcommand\Corr{\rho\rs{\theta N\rs{0}}}

\begin{document}

   \title{Investigating the accuracy achievable in reconstructing the angular sizes of stars through stellar intensity interferometry observations}
   \titlerunning{SII accuracy in reconstructing stellar angular sizes}


   \author{M. Fiori
          \inst{1}\fnmsep\inst{2}
          \and
          G. Naletto\inst{1}\fnmsep\inst{2}
          \and
          L. Zampieri\inst{2}
          \and
          I. Jiménez Martínez\inst{3}
          \and
          C. Wunderlich\inst{4}
          }

   \institute{Department of Physics and Astronomy, University of Padova, 
              Via F. Marzolo 8, I-35131, Padova, Italy\\
              \email{michele.fiori@unipd.it}
         \and
             INAF-Osservatorio Astronomico di Padova, 
             Vicolo dell'Osservatorio 5, I-35122, Padova, Italy
         \and
             Centro de Investigaciones Energéticas, Medioambientales y Tecnológicas,
             E-28040 Madrid, Spain
         \and
             University of Siena and INFN Pisa, 
             I-53100 Siena, Italy
             }

   \date{Received 23 May 2022 / Accepted 02 August 2022}

 
  \abstract
   {In recent years, stellar intensity interferometry has seen renewed interest from the astronomical community because it can be efficiently applied to Cherenkov telescope arrays.}
   {We have investigated the accuracy that can be achieved in reconstructing stellar sizes by fitting the visibility curve measured on the ground. The large number of expected available astronomical targets, the limited number of nights in a year, and the likely presence of multiple baselines will require careful planning of the observational strategy to maximise the scientific output.}
   {We studied the trend of the error on the estimated angular size, considering the uniform disk model, by varying several parameters related to the observations, such as the total number of measurements, the integration time, the signal-to-noise ratio, and different positions along the baseline.}
   {We found that measuring the value of the zero-baseline correlation is essential to obtain the best possible results. Systems that can measure this value directly or for which it is known in advance will have better sensitivity. We also found that to minimise the integration time, it is sufficient to obtain a second measurement at a baseline half-way between 0 and that corresponding to the first zero of the visibility function. This function does not have to be measured at multiple positions. Finally, we obtained some analytical expressions that can be used under specific conditions to determine the accuracy that can be achieved in reconstructing the angular size of a star in advance. This is useful to optimise the observation schedule.}
   {}


   \maketitle
%
\section{Introduction}

The accurate measurement of stellar sizes has always been of great relevance in astronomy.
Accurate knowledge of stellar radii is of fundamental importance for a comparison with various stellar evolution models that describe their physical properties \citep{Aufdenberg2005, Casagrande14}. Reaching angular resolutions of milliarcseconds, or even microarcseconds, is crucial to perform accurate measurements for many main-sequence and post-main-sequence stars and to foster other research areas, such as the study of fast-rotating stars, the implementation of accurate models for the limb-darkening effect (important for a correct derivation of star radii), the study of multiple star systems, and the investigation of the formation of hot spots on the surface of stars \citep{Monnier03, Labeyrie06, Barbieri2009, Dravins12}.

Measurements of stellar radii are obtained by means of interferometric techniques. The best-known technique is the (phase) interferometry (e.g. Michelson interferometer, \citealt{Pease1931}), which measures the first-order spatial correlation of the radiation emitted by a source. A less well-known technique is intensity interferometry, which exploits the measurement of the second-order spatial correlation of the radiation from a star \citep{Glauber63}. This technique, referred to as stellar intensity interferometry (SII), found an astronomical application in the optical band at the Narrabri Stellar Intensity Interferometer (NSII) in the 1960s through the pioneering experiments of Hanbury Brown and Twiss \citep{HBT54, HBT56, HBT57, HBT58, HBT74a}. 
For several decades, this technique was almost forgotten and has come back to life through newly developed technologies and the combined efforts of a number of collaborations \citep{Naletto2016, Zampieri16, Tan2016, Guerin17, Weiss2018, Matthews2018, Rivet20, MAGIC19, VERITAS20, Zampieri2021, Fiori2021, Horch2021}. The main efforts today are made in the development of hardware and software that are suitable for implementing this technique on existing Cherenkov telescopes  \citep{Matthews2018, MAGIC19, VERITAS20} as well as on future arrays \citep{Dravins2013, Scuderi21, Vercellone22}. The large collecting area and the fast optics of this type of telescopes enable studying stars that are much dimmer than those observed with the NSII \citep{Rou2013}.\\

In this work, we investigate the accuracy that can be achieved in the measurement of stellar radii by considering one or more intensity interferometry observations. Our goal is to derive relations that can describe the error trend on the final fitted radius and provide simple prescriptions to optimise the data-taking process. To do this, we studied the trend of the error on the estimated angular size by varying several parameters related to the observations, such as the total number of measurements, the exposure time, the signal-to-noise ratio, and different positions along the baseline. We used the uniform disk model for the radial profile of the stars, which was fitted to the data using a least-squares method, which provides the associated error measurement.
Despite the simplified approach, this study is definitely of interest for the proper planning of SII observations with Cherenkov telescopes. The large number of available astronomical targets, the limited number of nights in a year (generally, SII observations will be restricted to nights around full moon because Cherenkov observations are not possible during these periods), and the likely presence of multiple baselines will require careful planning to maximise the scientific outcome. Deriving a few simple prescriptions, together with a quick way to determine the integration time needed to reach a certain accuracy, is very useful to this end.\\

The paper is structured as follows. In Sect. \ref{sec:method} we briefly describe the SII background and the method we used to obtain the information on the errors from the fits. In the following sections, Sects. \ref{sec:onepoint} and \ref{sec:twopoints}, we show the results of our analyses, and in Sect. \ref{sec:simuluations}, we show a simulation we conducted to probe the validity of our results. Finally, in Sect. \ref{sec:conclusion} we briefly summarise our results and draw the conclusions.

\section{Method}
\label{sec:method}

Stellar intensity interferometry is based on the fact that the intensities of the signal coming from a thermal source and measured with two telescopes on the ground are correlated up to a certain degree. We can define the second-order coherence function $g^{(2)}$ as follows:
\begin{equation}
    g^{(2)}(\tau, d) = \frac{<I\rs{A}(t, d) \cdot I\rs{B}(t+\tau, d) >}{<I\rs{A}(t)>\cdot <I\rs{B}(t)>},
\end{equation}
where $<\;>$ denotes the average over time, and $I\rs{A}(t, d)$ and $I\rs{B}(t+\tau, d)$ are the light intensities recorded at a certain distance $d$ and at a certain time $t$, accounting for the light travel-time delay $\tau$ between the two detectors. For $\tau=0,$ the relation between the second-order coherence function and the squared visibility function $|\Vis(d)|^2$, which for a distant observer, is the Fourier transform of the source brightness distribution of the star, can be written as follows:
\begin{equation}\label{eq:g2}
    g^{(2)}(0, d) = 1+ N\rs{0}|\Vis(d)|^2,
\end{equation}
where $N\rs{0}$ is a normalisation factor that depends on the approach that is used to measure the correlation (e.g. photon-counting versus continuous mode) and on the properties of the observation system.  
In the continuous mode, which is the original SII method implemented by Hanbury Brown and Twiss \citep{HBT74a}, the light intensities at the two detectors are converted into currents and are continuously recorded by means of acquisition systems with electronic bandwidth $\Delta f$ (the reciprocal of the sampling time of the radiation intensities). This bandwidth is generally much smaller than the optical bandwidth $\Delta \nu$ (the reciprocal of the radiation coherence time), and therefore $N\rs{0}$ is equal to the ratio of the two bandwidths,
\begin{equation}
N\rs{0} = \frac{\Delta f}{\Delta\nu}.   
\end{equation}
In the case of the photon-counting technique, where the arrival time of photons at the two detectors is recorded and then correlated, we can make a similar argument considering that the coherence time of light $\tau\rs{c}$ is generally much shorter than the sampling time $dt$ (for more details, see \citealt{Naletto2016}),  
\begin{equation}
N\rs{0} = \frac{\tau\rs{c}}{dt}.   
\end{equation}

The visibility function of a star on the ground can be approximated to first order with a uniform disk,
\begin{equation}\label{eq:uniformdisk}
    \VisSq = \left|2 \frac{J\rs{1}(\pi \theta d / \lambda)}{\pi \theta d / \lambda}\right|^2,
\end{equation}
where $J\rs{1}$ is the Bessel function of the first order, $\lambda$ is the central wavelength of the filter used for the observations, $d$ is the separation between the two telescopes, and $\theta$ is the angular size of the star.
For the purpose of this work, that is, to study the trend of the errors on the estimated stellar sizes, this model is fairly accurate because our analysis is limited only to the first peak of $\VisSq$ where other effects (such as the limb-darkening effect and/or hot spots on the stellar surface) are less important \citep{Berger2007}. Moreover, a series of correction factors have been reported that can be used to convert the uniform-disk diameter into a more general limb-darkened diameter considering the type of the observed star \citep[see e.g. ][]{HanburyBrownLimbDarkening1974}.

For sake of simplicity, we directly consider the function $\Gamma = g^{(2)}-1$, without considering the method that was selected to obtain the measurements. We then focus on the analysis of the visibility curve, Eq. \eqref{eq:uniformdisk}, including the normalisation factor $N\rs{0}$ (by means of Eq. \ref{eq:g2}), that is,
\begin{equation}\label{eq:UDnorm}
\Gamma = N\rs{0}\VisSq = N\rs{0}\left|2 \frac{J\rs{1}(\pi \theta d / \lambda)}{\pi \theta d / \lambda}\right|^2.
\end{equation}
In addition, for convenience, hereafter we call the parameter $N\rs{0}$ the zero-baseline correlation (ZBC) value because this is the value of the visibility curve at a baseline equal to zero.
These equations are valid for both approaches, but in the photon-counting case, the ZBC value is generally known (or it is possible to measure it precisely), and Eq. \eqref{eq:uniformdisk} can be directly fitted to the data (see e.g. \citealt{Zampieri2021}, where the measured ZBC value agrees with expectations). 
To study the errors of the fitted angular sizes, we generated one or more measurements at different baselines and with different uncertainties, following equation \eqref{eq:UDnorm} (or \eqref{eq:uniformdisk} for the case with known ZBC). During the generation of the different models for the different scenarios, random values were set for the parameters $N\rs{0}$, $\theta$ , and $\lambda$ from a uniform distribution ($N\rs{0} \in [10^{-4}:10^{-1}]$, $\theta \in [0.4:2.0]$ mas, and $\lambda \in [400:500]$ nm) to avoid any kind of bias and to show that the results are valid for any value of these parameters. The simulated measurements were then fitted again with the same equation as was used to generate them, leaving as free parameters $\theta$ and $N\rs{0}$ (or only $\theta$ in the case with known ZBC). To fit the curves, we used a least-squares minimisation algorithm that allowed us to estimate the covariance matrix and extract the errors on the fitted parameters. We then finally carried out a full simulation to compare the results with a realistic situation.

\begin{figure}
    \centering
    \includegraphics[width=1.\columnwidth]{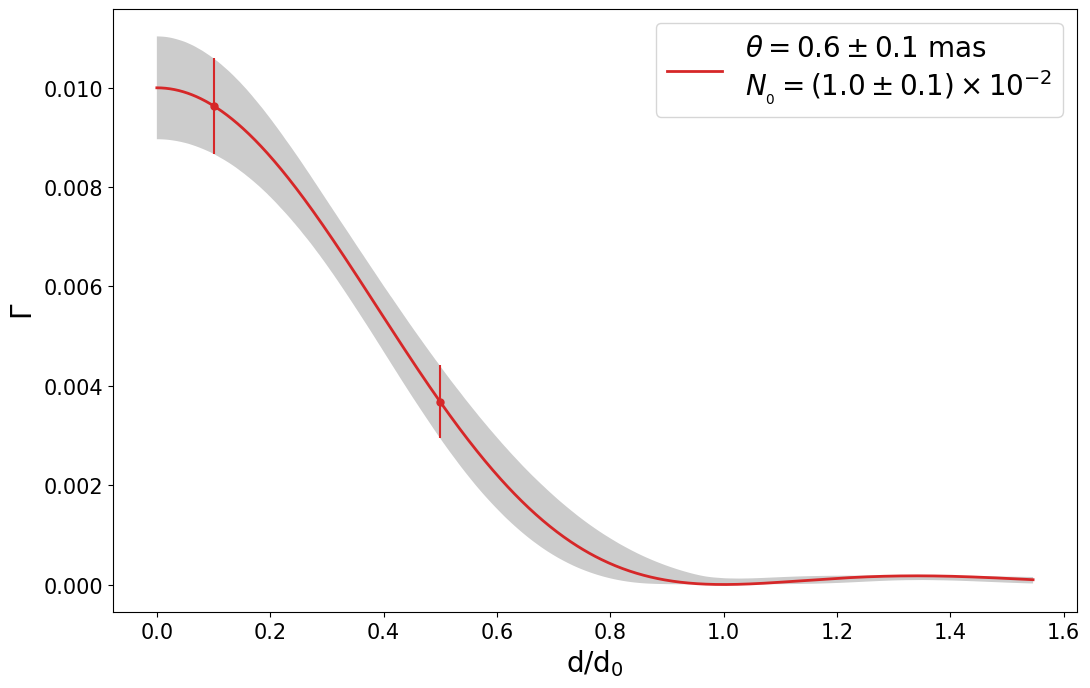}
    \caption{Example of the result of the fitting procedure we used to compute the error on $\theta$ in the following sections. We have considered a star with an angular size $\theta=0.6$ mas and a ZBC value of $N\rs{0}=10^{-2}$. The grey shaded area correspond to the 1$\sigma$ confidence interval.}
    \label{fig:fit_example}
\end{figure}

In Fig. \ref{fig:fit_example} we show an example of the typical fit in this work, considering a star with a radius $\theta=0.6$ mas and a ZBC value of $N\rs{0}=10^{-2}$. We generated two measurements of the visibility curve with a certain error and fitted Eq. \eqref{eq:UDnorm} on them. From this fit, we then extracted the error associated with the reconstructed stellar angular size $\theta$.
The baselines in the plot are renormalised to the value of the baseline $d\rs{0}$ , corresponding to the first zero of the first-order Bessel function $J\rs{1}$ (i.e. the baseline of the first peak of the visibility function goes from 0 to 1). We  refer to the normalised baseline $d/d\rs{0}$ throughout when we talk about the measurement positions on the visibility curve.\\

A widely used parameter to quantify the significance of a measurement is the signal-to-noise ratio (S/N).
Operationally, the S/N is computed taking a measured value and dividing it by the associated uncertainty.
For the measurements on the visibility curve $\VisSq$, the theoretical value of the S/N was computed by \cite{HBT74a} and is (in the case of unpolarized light)
\begin{equation}\label{eq:SNgam}
\SNgam = N \beta \VisSq\sqrt{T/2} = \alpha \VisSq \sqrt{T},
\end{equation}
where $N$ is a term that depends on the source photon rate (and can be expressed as a count-rate or as a flux density multiplied by a collection area), $\beta$ is a term that depends on the system performances (detector quantum efficiency, optical bandwidth, or coherence time of the light, electronic bandwidth, or sampling time, etc.), and $T$ is the total integration time. For our purposes, we can consider the term depending on the source photon rate as a constant, and we incorporated it together with the other constants in the term $\alpha$, which were considered equal to 1. We show at the end, with the full simulation performed in Sect. \ref{sec:simuluations}, that the results are valid for any value of $\alpha$.

By analogy, we can work in terms of S/N also in the case of the final measurements of the stellar size. Instead of speaking of errors on $\theta,$ we consider $\SNtheta$. This is useful to compare the results in different situations.
We show that we can define some analytical expressions that describe the trends of $\SNtheta$ as a function of $\SNgam$ or of the integration time.

\section{Simplest case: Known ZBC}
\label{sec:onepoint}

\begin{figure}
    \centering
    \includegraphics[width=1\columnwidth]{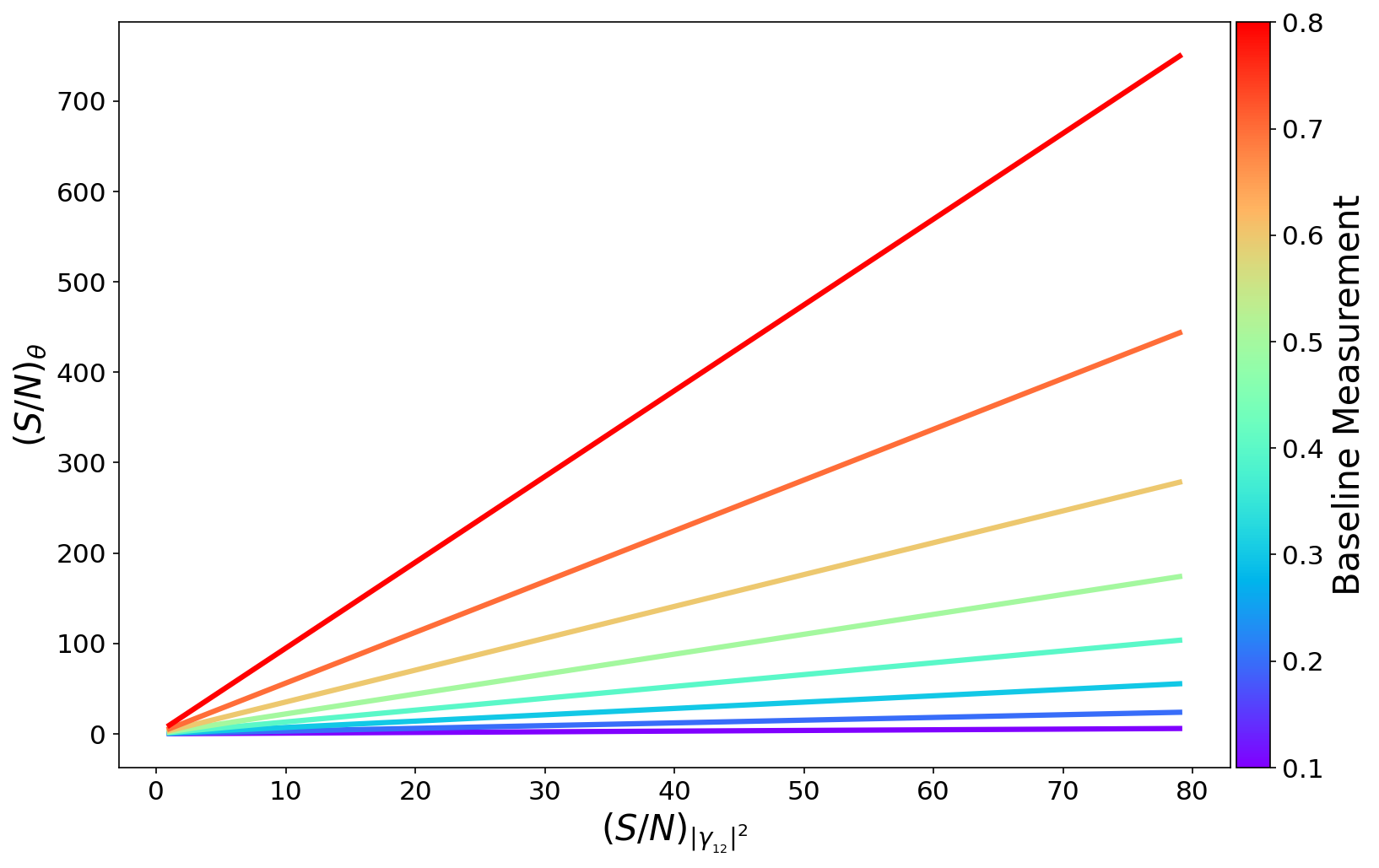}
    \includegraphics[width=1\columnwidth]{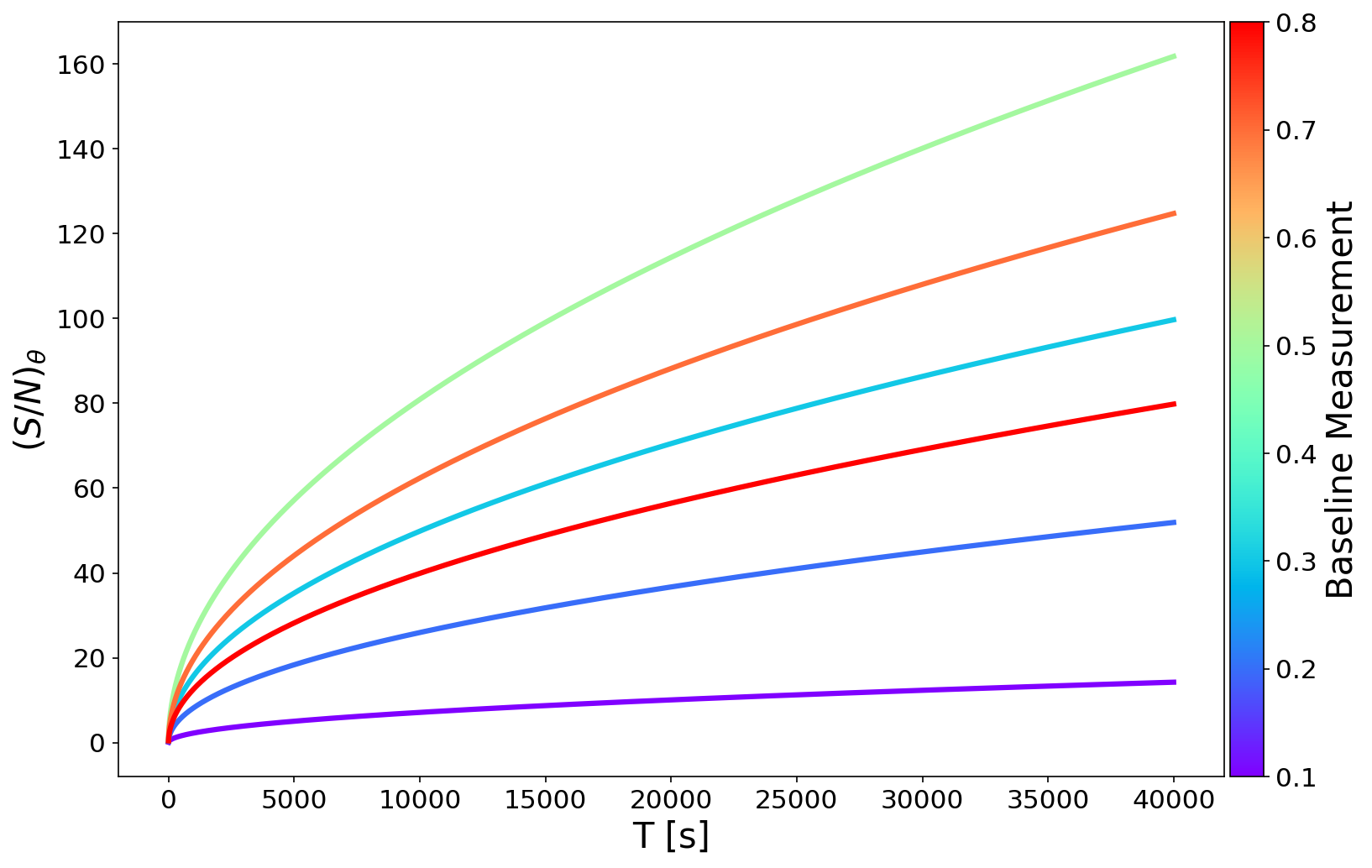}
    \caption{Trend of $\SNtheta$ as a function of $\SNgam$ (upper panel) and of the integration time (lower panel) inferred from the simulations for measurements at different normalised baselines (differently coloured curves).}
    \label{fig:1p_SN_T}
\end{figure}

We start by considering the simplest case, that is, when the ZBC value is known. 
In this situation, we need a single measurement almost anywhere along the visibility curve in order to constrain the model (Eq. \ref{eq:uniformdisk}).
Therefore, we studied the behaviour of $\SNtheta$ by changing the $\SNgam$ or the integration time $T$ of this single measurement. This analysis was limited to the interval $d/d\rs{0}=[0.0,0.8]$ because using a measurement at a normalised baseline larger than 0.8 does not allow us to unambiguously fit the size of the star (the measurement could belong to either the first peak or the second peak of the visibility curve) without imposing constraints during the fitting procedure. The limit at 0.8 was also kept fixed for all the other analyses in this work.  

\begin{figure}
    \centering
    \includegraphics[width=1\columnwidth]{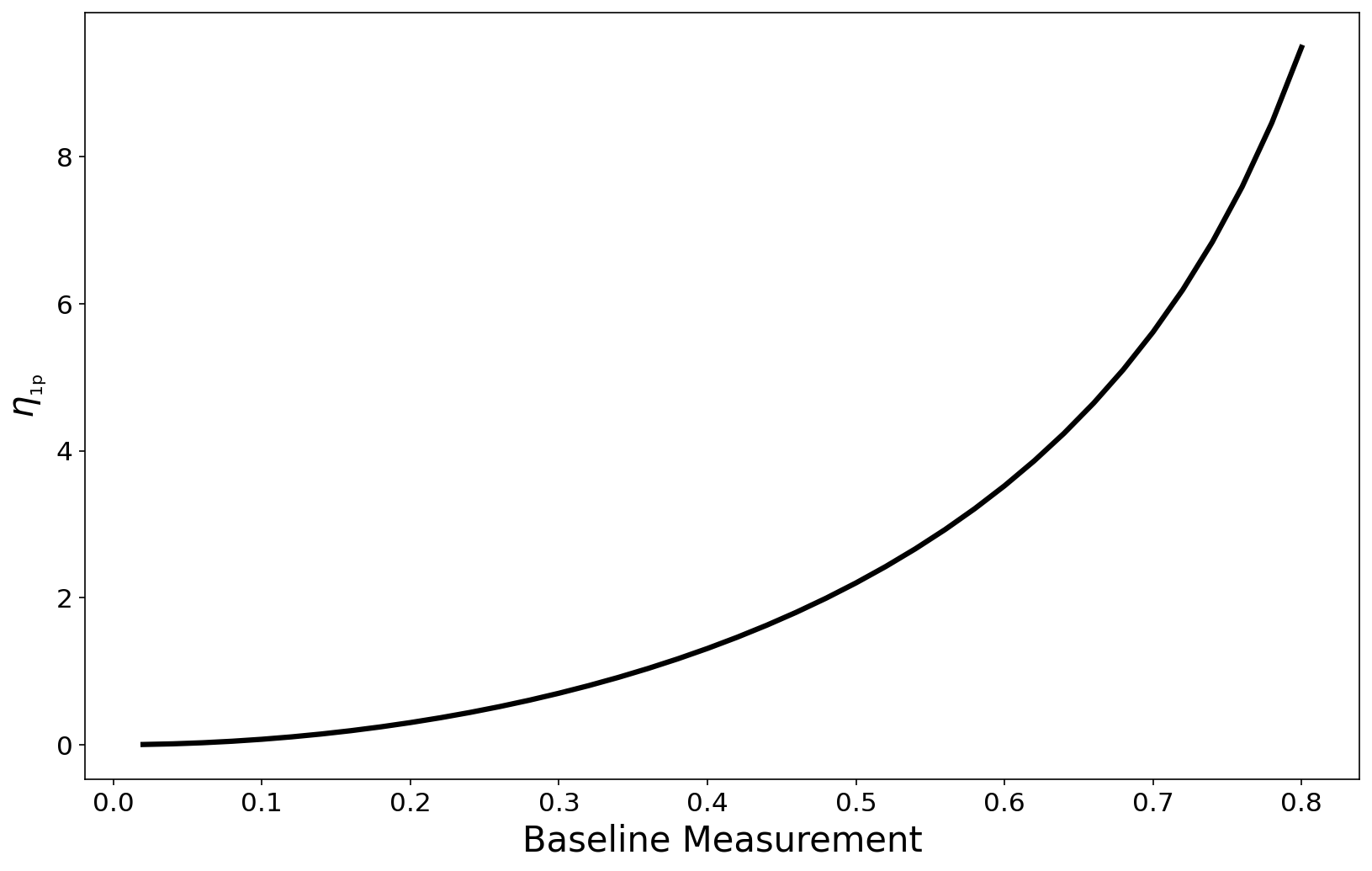}
    \includegraphics[width=1\columnwidth]{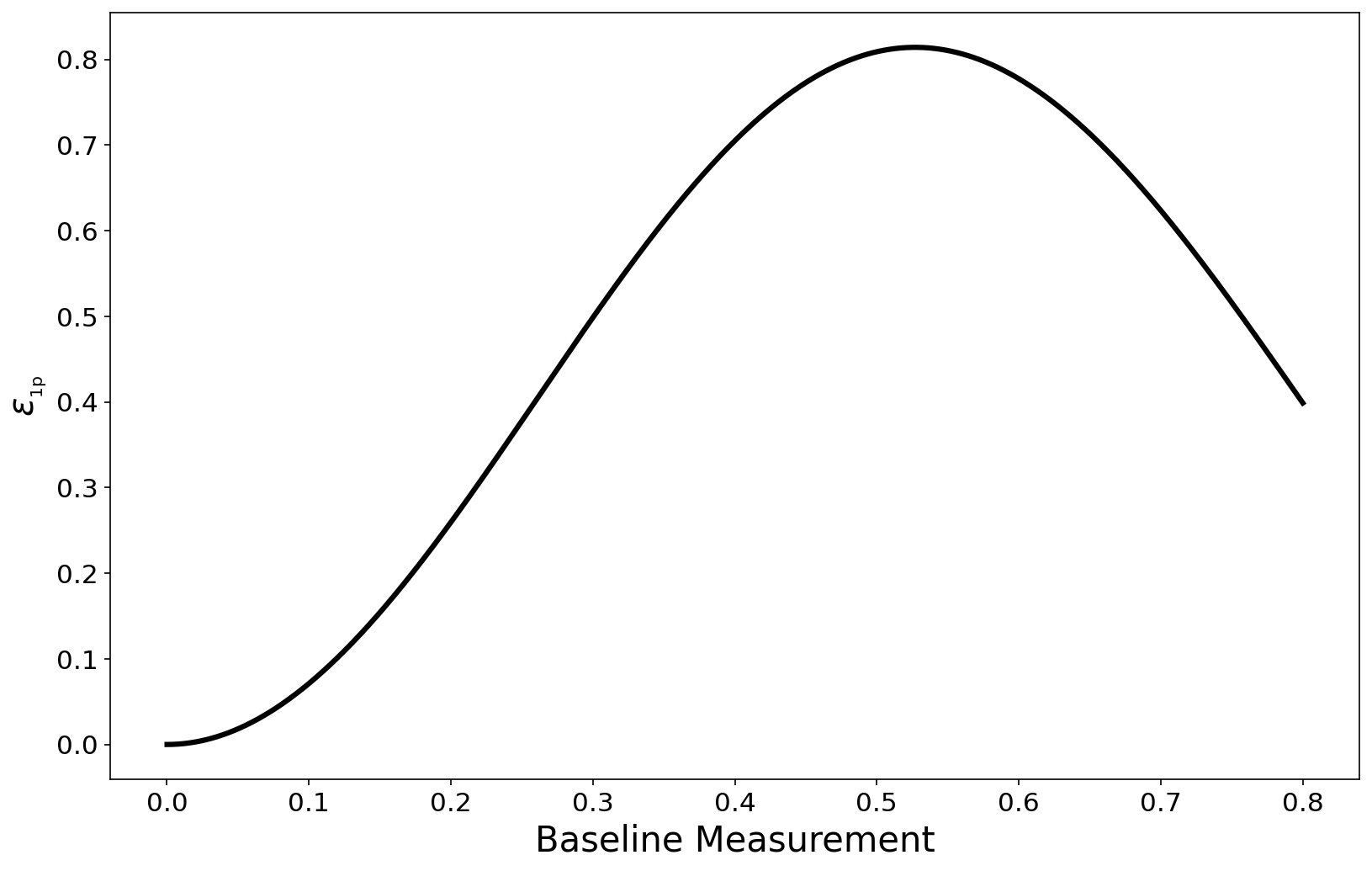}
    \caption{Trends of  $\etp$ (upper panel) and $\epsp$ (lower panel) coefficients as functions of the baseline of the measurement.}
    \label{fig:1p_eta_eps}
\end{figure}

In Fig. \ref{fig:1p_SN_T} we show the trend of $\SNtheta$ as a function of $\SNgam$ (upper panel) or of the integration time $T$ (lower panel) for different baselines (differently coloured curves). The curves were obtained by extracting the error associated with the reconstructed angular size, as discussed above, from a series of fits in which we changed the parameters.
We infer a linear relation between $\SNtheta$ and $\SNgam$ and a square-root relation between $\SNtheta$ and $T$ (similarly to $\SNgam\propto\sqrt{T}$).In this simple situation, Eq. \eqref{eq:etp} might appear to be redundant because we can directly expect $\SNtheta\propto\SNgam\propto\sqrt{T}$. However, later in this work, this relation is no longer valid, and it will be useful to have explicitly found an expression for $\etp$. This suggests that we can write the two following equations,
\begin{subequations}
    \begin{equation}\label{eq:etp}
        \SNtheta = \etp\SNgam,
    \end{equation} 
    \begin{equation}\label{eq:epsp}
        \SNtheta = \epsp\sqrt{T},
    \end{equation}    
\end{subequations}
where we introduced the proportionality parameters $\etp$ and $\epsp$.
In this simple situation, Eq. \eqref{eq:etp} might appear to be redundant because we can directly expect $\SNtheta\propto\SNgam\propto\sqrt{T}$. However, later in this work, this relation is no longer valid, and it will be useful to have explicitly found an expression for $\etp$.
The top panel of Fig. \ref{fig:1p_SN_T} shows that by increasing $\SNgam$, $\SNtheta$ always increases linearly, and that the larger the baseline of the measurement, the higher the $\SNtheta$ value. 
The lower panel shows that $\SNtheta$ grows proportionally to the square root of the integration time, and as the baseline of the measurement increases, the $\SNtheta$ value (for a given $T$) increases up to a certain baseline ($\sim0.5$). For baselines larger than 0.5--0.6, the S/N tends to decrease. This fact can be explained by considering that at longer normalised baselines, $\SNtheta$ improves (as shown in the upper panel) and the integration time needed to achieve a certain $\SNgam$ always increases as the normalised baseline increases. At a given point, the integration time is no longer sufficient to compensate for the reduction of the S/N caused by the decrease in $\VisSq$ as the normalised baseline increase.
Figure \ref{fig:1p_eta_eps} shows a different representation of this behaviour: $\etp$ (upper panel) grows monotonically with the baseline, while $\epsp$ (lower panel) grows up to a maximum value and then decreases. 
It is possible to provide analytical expressions of the parameters $\etp$ and $\epsp$. From the error propagation of Eq. \eqref{eq:uniformdisk}, assuming that $d$ and $\lambda$ are perfectly known (or that the associated errors are much smaller than the error on the stellar size $\theta$), we obtain
\begin{equation}\label{eq:errprop_1p}
    \sigma^2\rs{\VisSq} = \biggl(\frac{\partial \VisSq}{\partial \theta}\biggr)^2\sigma^2\rs{\theta},
\end{equation}
where $\sigma\rs{\VisSq}$ is the error associated with the measurement on the visibility curve ($\VisSq$) and $\sigma\rs{\theta}$ is the error associated with the angular size of the star. Inverting Eq. \eqref{eq:etp}, we find
\begin{equation}\label{eq:sigmatheta_eta}
    \sigma\rs{\theta} = \frac{\theta}{\VisSq}\frac{\sigma\rs{\VisSq}}{\etp},
\end{equation}
where we considered $\SNtheta = \theta/\sigma\rs{\theta}$ and $\SNgam = \VisSq/\sigma\rs{\VisSq}$. Computing the derivative of Eq. \eqref{eq:uniformdisk} for $\theta$, we find
\begin{equation}
   \frac{\partial\VisSq}{\partial\theta} = -\frac{8J\rs{1}(\pi \theta d / \lambda) J\rs{2}(\pi \theta d / \lambda)}{\pi \theta^2 d / \lambda},
\end{equation}
and inserting this equation, together with Eq. \eqref{eq:sigmatheta_eta}, in Eq. \eqref{eq:errprop_1p} we obtain
\begin{equation}\label{eq:etpcoeff}
    \etp = \biggl|\frac{2J\rs{2}(\pi \theta d / \lambda)}{J\rs{1}(\pi \theta d / \lambda)}(\pi \theta d / \lambda)\biggr|.
\end{equation}

We can do a similar exercise for $\epsp$. In this case, we consider Eq. \eqref{eq:epsp} and \eqref{eq:SNgam} (with $\alpha=1$) to find
\begin{subequations}
    \begin{equation}\label{eq:sigmatheta_eps}
        \sigma\rs{\theta} = \frac{\theta}{\epsp}\frac{1}{\sqrt{T}},
    \end{equation}
    \begin{equation}\label{eq:sigmaVis_eps}
        \sigma\rs{\VisSq} = \frac{1}{\sqrt{T}},
    \end{equation}        
\end{subequations}
and, inserting them in Eq. \eqref{eq:errprop_1p}, we obtain
\begin{equation}\label{eq:epspcoeff}
    \epsp = \biggl|\frac{8J\rs{1}(\pi \theta d / \lambda)J\rs{2}(\pi \theta d / \lambda)}{(\pi \theta d / \lambda)}\biggr|.
\end{equation}
As expected from eqs. \eqref{eq:SNgam}, \eqref{eq:etp} and \eqref{eq:epsp}, we find that $\epsp = \VisSq\etp$.
From this analytical solution for $\epsp$ , we find that the maximum $\SNtheta$ is reached at a normalised baseline $d/d\rs{0} \simeq 0.527$. This same value can also easily be found numerically. This means that in this scenario, this is the best position along the visibility curve to maximise $\SNtheta$. 

\section{General case: Unknown ZBC}
\label{sec:twopoints}
In a more general scenario, the ZBC value is not known in advance and needs to be fitted together with the visibility curve using Eq. \eqref{eq:UDnorm}. To obtain a measurement of the stellar size, we need at least two measurements on the visibility curve.

In these conditions, the analysis is slightly more elaborate, as we can have two or more different S/Ns or two or more different integration times for the measurements on the visibility curve. 
We first consider the case in which there are only two measurements with the same S/N or the same integration time because these two conditions are not unlikely. For example, many future SII Cherenkov telescopes will be able to carry out observations simultaneously on both long and short baselines (ZBC value), so that in the end, the integration time will always be the same. Even the case in which the S/Ns are similar is not unlikely because observational strategies might be designed in this way. However, we also discuss how the results change when the S/N or the integration time are different and what happens when an additional measurement is added to the analysis.

\subsection{Two measurements with the same S/N or integration time}

\begin{figure}
    \centering
    \includegraphics[width=1\columnwidth]{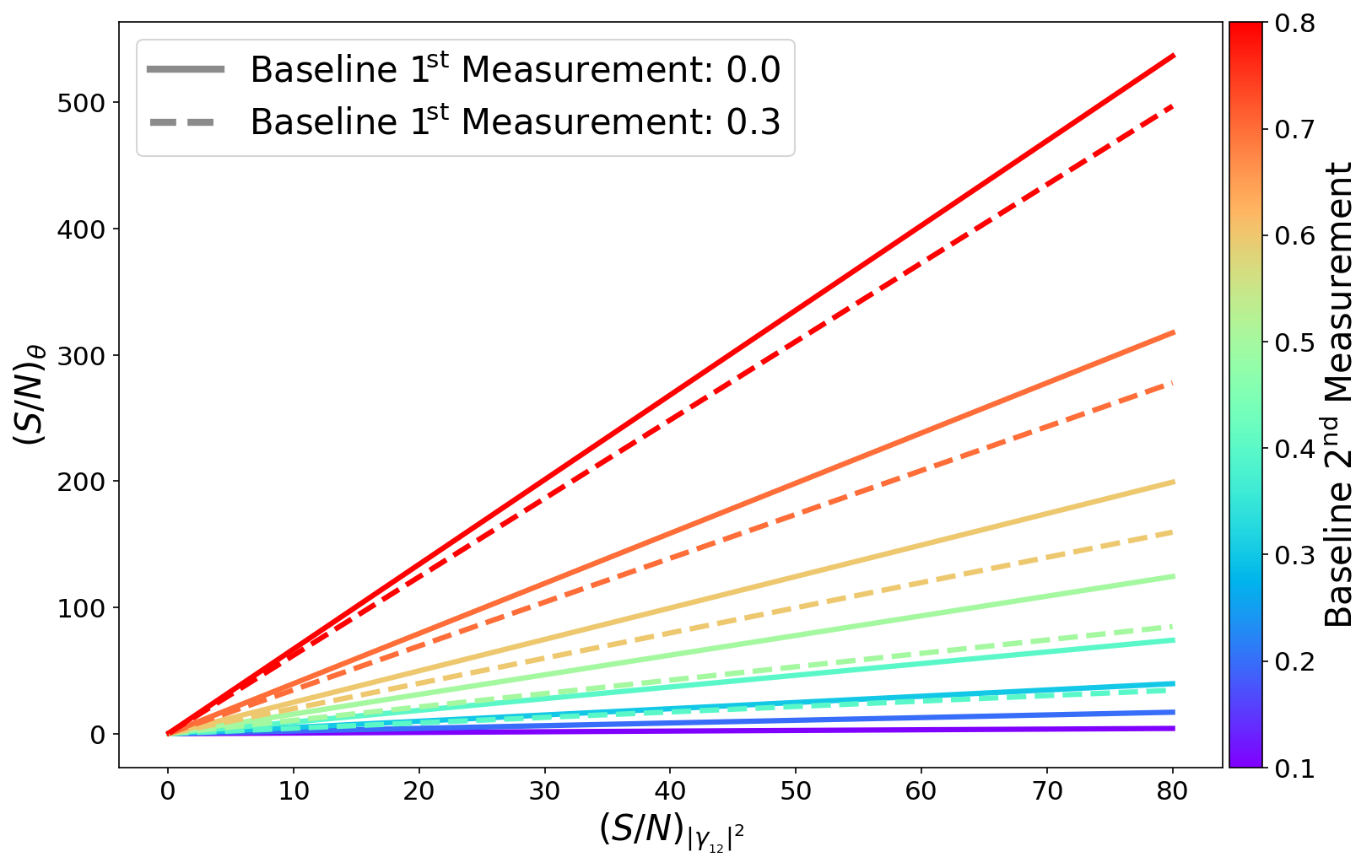}
    \includegraphics[width=1\columnwidth]{ 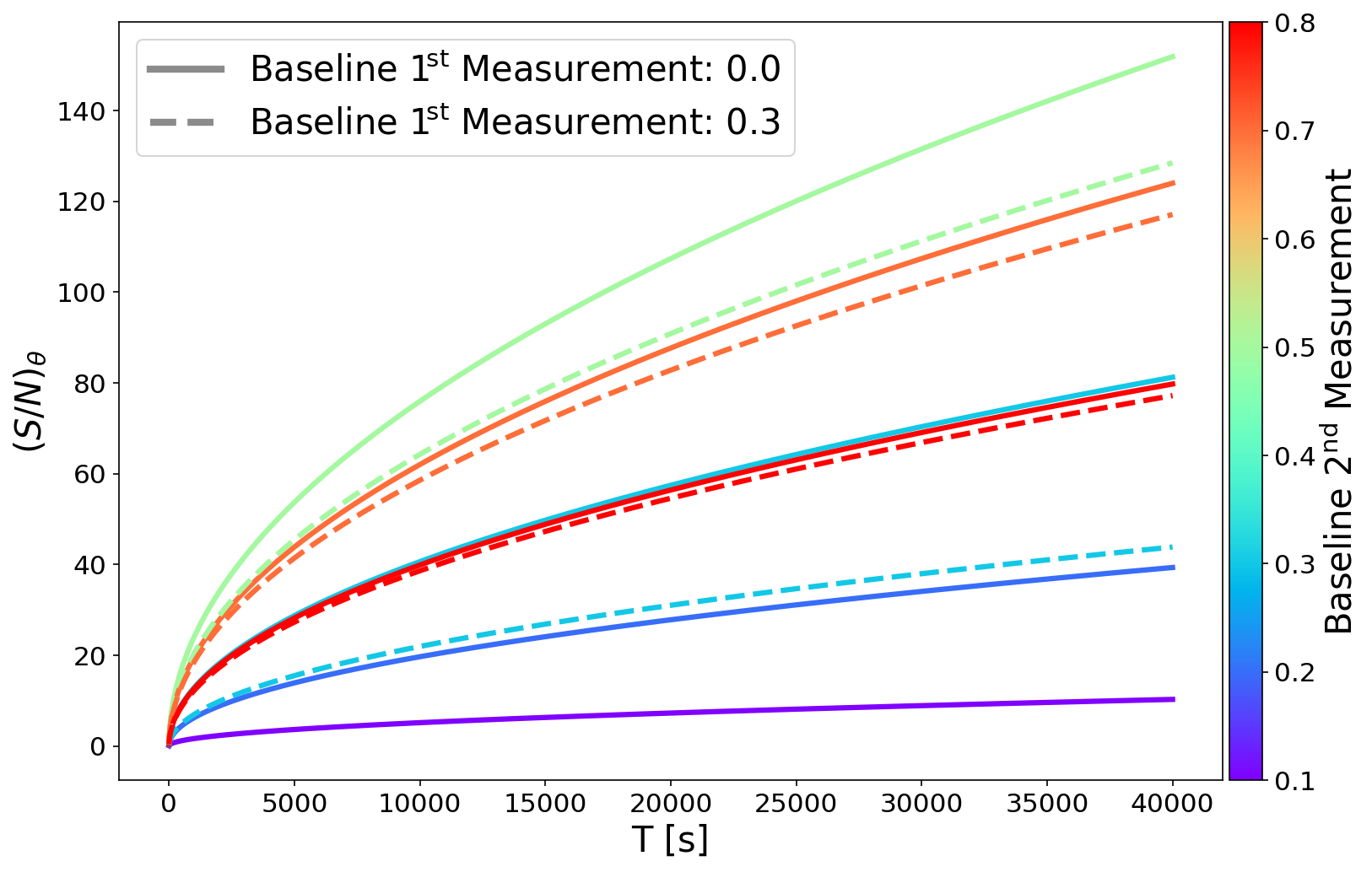}
    \caption{Trend of $\SNtheta$ as a function of $\SNgam$ (upper panel) and of the integration time (lower panel). Solid curves correspond to the cases in which the normalised baseline of the first measurement is at $d\rs{1}/d\rs{0} = 0.0$, and dashed curves correspond to the cases in which the baseline of the first measurement is at $d\rs{1}/d\rs{0} = 0.3$ (value chosen to ensure a clear separation between the dashed and continuous curves). The differently coloured curves depend on the normalised baseline of the second measurement.}
    \label{fig:2p_SN_T}
\end{figure}

In this configuration, the results are similar to those seen in the previous section. 
Results are shown in Fig. \ref{fig:2p_SN_T}, where we show the trend of $\SNtheta$ as a function of $\SNgam$ in the upper plot and as a function of the integration time in the lower plot. 
The trend of $\SNtheta$ as a function of $\SNgam$ is linear, while $\SNtheta$ varies with the square root of the integration time. The main difference with the previous case is that we always find a value for $\SNtheta$ lower than that for the case where the ZBC is known (because now there is an additional parameter in the fitting procedure). 
It is now useful to note that as the position of the lower baseline measurement increases, the maximum value of $\SNtheta$ tends to decrease. The lower plot also shows that by increasing the position of the first measurement, the maximum values of $\SNtheta$ are obtained for longer normalised baselines of the second measurement.

\begin{figure}
    \centering
    \includegraphics[width=1\columnwidth]{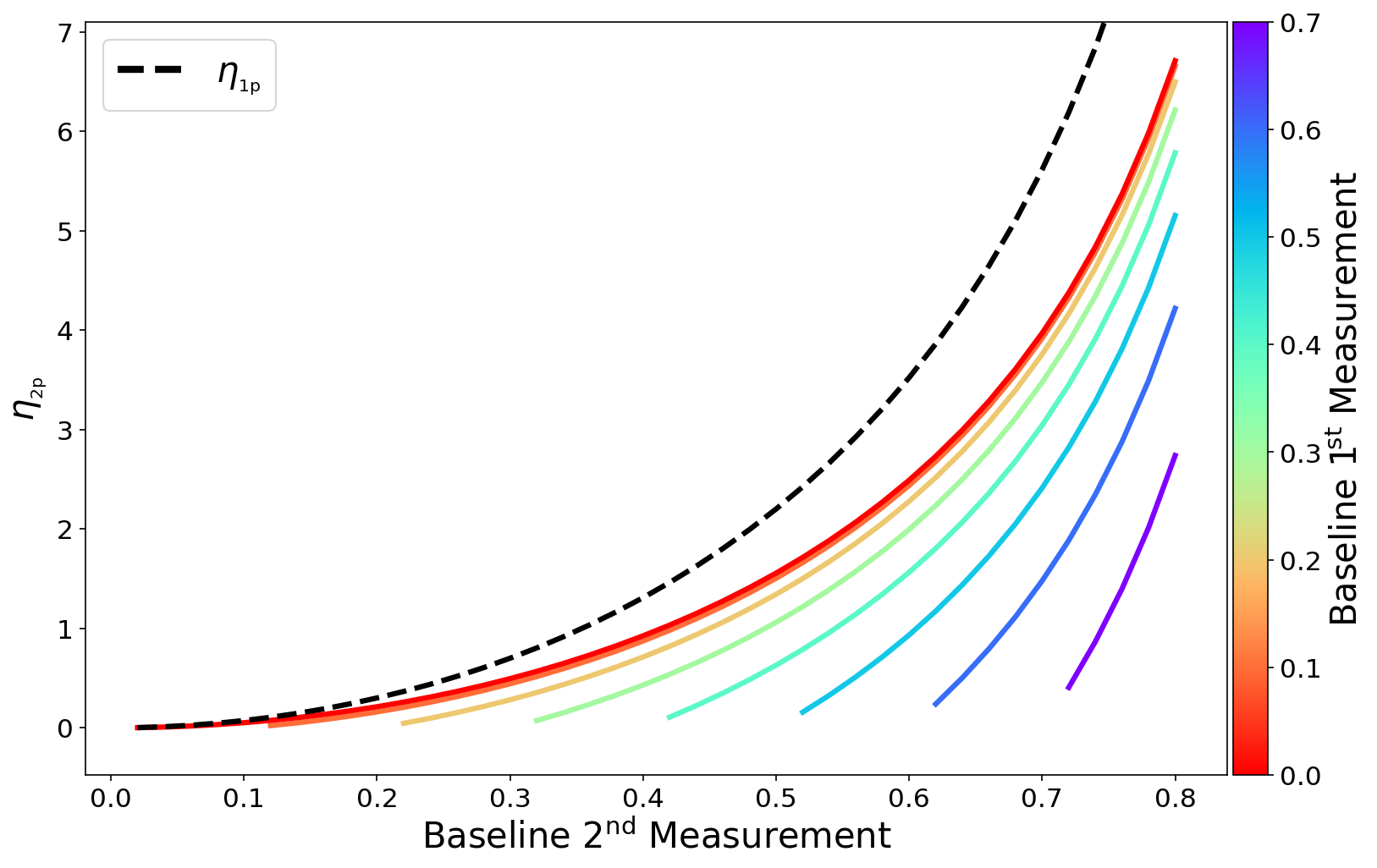}
    \includegraphics[width=1\columnwidth]{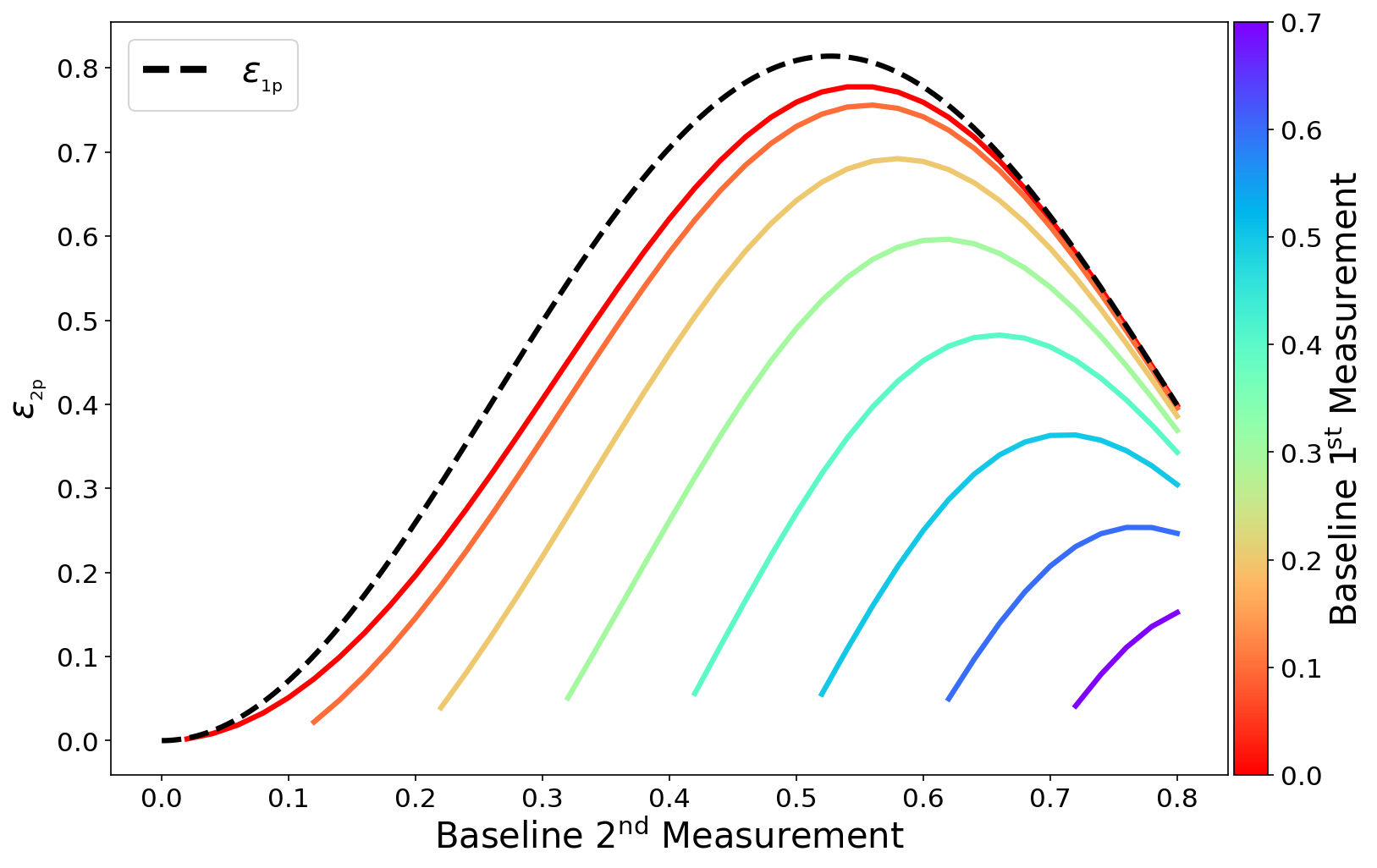}
    \caption{Trend of  $\etpp$ (upper panel) and $\epspp$ (lower panel) coefficients as function of the baselines of the first measurement (differently coloured curves) and of the second measurement ($x$-axis). We show as dashed black curves the trends of $\etp$ and $\epsp$ (already shown in Fig. \ref{fig:1p_eta_eps}) for comparison. The redder curves in the plots correspond to the trends of $\etzbc$ and $\epszbc$, when the ZBC value is directly measured.}
    \label{fig:etaeps_2p}
\end{figure}

We can thus again define two parametrical coefficients, $\etpp$ and $\epspp$, that link $\SNtheta$ with $\SNgam$ and $T$, as done in the previous section,
\begin{subequations}
    \begin{equation}\label{eq:etpp}
        \SNtheta = \etpp\SNgam,
    \end{equation} 
    \begin{equation}\label{eq:epspp}
        \SNtheta = \epspp\sqrt{T}.
    \end{equation}    
\end{subequations}
The trends of the two coefficients as function of the baselines of the measurements are shown in Fig. \ref{fig:etaeps_2p} (upper panel for $\etpp$ and lower panel for $\epspp$). In the plots we also show the curves of $\etp$ and $\epsp$ (dashed black lines) computed above. The shapes of the curves are similar to the shape of the curve for the known ZBC scenario, but the maximum value continually decreases (as the baseline of the first measurement is increased).

The similarity of these curves with those in the known ZBC scenario suggests that we can find an analytical expression for $\etpp$ and $\epspp$. This is definitely true at least for the case where the shortest baseline measurement is at $d/d\rs{0} = 0$, when we measure the ZBC value directly. 
We start from the error propagation formula for $\Gamma$ (Eq. \ref{eq:UDnorm}) to obtain the expressions for $\etzbc$ and $\epszbc$ (which are particular cases of $\etpp$ and $\epspp$ when the shortest measurement directly provides the ZBC).
The error propagation formula is slightly more complex now because there are two parameters that can also be correlated,
\begin{equation}\label{eq:errprop_2p}
    \sigma^2\rs{\Gamma} = \biggl(\frac{\partial\Gamma}{\partial\theta}\biggr)^2\sigma\rs{\theta}^2 + 
                          \biggl(\frac{\partial\Gamma}{\partial N\rs{0}}\biggr)^2\sigma\rs{N\rs{0}}^2 +
                          2\biggl(\frac{\partial\Gamma}{\partial\theta}\biggr)\biggl(\frac{\partial\Gamma}{\partial N\rs{0}}\biggr)\sigma\rs{\theta}\sigma\rs{N\rs{0}}\Corr,
\end{equation}
where $\Corr$ is the correlation coefficient between $\theta$ and $N\rs{0}$.
Making the derivatives of $\Gamma$ with respect to $\theta$ and $N\rs{0}$ , we obtain 
\begin{subequations}
    \begin{equation}\label{eq:der_GamTheta}
    \frac{\partial\Gamma}{\partial\theta} = -N\rs{0} \frac{8J\rs{1}(\pi \theta d / \lambda) J\rs{2}(\pi \theta d / \lambda)}{\pi \theta^2 d / \lambda},
    \end{equation} 
    \begin{equation}\label{eq:der_GamN0}
    \frac{\partial\Gamma}{\partial N\rs{0}} = \left|2 \frac{J\rs{1}(\pi \theta d / \lambda)}{\pi \theta d / \lambda}\right|^2 
    = \frac{\Gamma}{N\rs{0}}.
    \end{equation}    
\end{subequations}
In order to find $\etzbc$ , we can simplify some terms. First of all, we can consider the case in which the two measurements have the same $\SNgam$ and the error on the measurement at $d/d\rs{0} = 0$ corresponds to the error on the ZBC value\footnote{Meaning that $\Gamma\rs{1}/\sigma\rs{\Gamma\rs{1}} = \Gamma\rs{2}/\sigma\rs{\Gamma\rs{2}}$ (where the subscripts 1 stands for the lower baseline measurement and 2 for the higher ones) and that $N\rs{0} \equiv \Gamma\rs{1}$ and $\sigma\rs{N\rs{0}} \equiv \sigma\rs{\Gamma\rs{1}}$.}, 
\begin{equation}
    \sigma\rs{N\rs{0}} = \frac{N\rs{0}}{\Gamma}\sigma\rs{\Gamma}.
\end{equation}
Then we again consider Eq. \eqref{eq:sigmatheta_eta}, replacing $\etp$ with $\etzbc$.
Inserting these equations in Eq. \eqref{eq:errprop_2p}, after some algebra, we obtain
\begin{equation}\label{eq:etazbcwithrho}
    \etzbc = \biggl|\frac{2J\rs{2}(\pi \theta d\rs{2} / \lambda)}{J\rs{1}(\pi \theta d\rs{2} / \lambda)} (\pi \theta d\rs{2} / \lambda)\biggr|
             \frac{1}{2\Corr},
\end{equation}
where we have made explicit that the results depend on the baseline of the second point.
Comparing Eq. \eqref{eq:etazbcwithrho} with Eq. \eqref{eq:etpcoeff}, we can see that $\etzbc = \etp/(2\Corr)$. The value of $\Corr$ can then be calculated from the numerical values of $\etp$ and $\etzbc$ computed from the simulations (the black dashed curve and the redder continuous curve in the upper panel of Fig. \ref{fig:etaeps_2p}). We find that
\begin{equation}\label{eq:corr_etzbc}
     \Corr = \frac{\sqrt{2}}{{2}},
\end{equation}
meaning a fairly high correlation between the two parameters. The final expression for $\etzbc$ is
\begin{equation}\label{eq:etzbc}
    \etzbc = \biggl|\frac{\sqrt{2}J\rs{2}(\pi \theta d\rs{2} / \lambda)}{J\rs{1}(\pi \theta d\rs{2} / \lambda)} (\pi \theta d\rs{2} / \lambda)\biggr|.
\end{equation}\\

Similarly, to determine $\epszbc$, we simplified further. We used eq. \eqref{eq:sigmatheta_eps} (replacing $\epszbc$ with $\epsp$) and considered that the error of the two measurements is the same\footnote{Since $(S/N)\rs{\Gamma_i} \equiv \Gamma_i/\sigma\rs{\Gamma_i} = \Gamma_i\sqrt{T}$ with equal exposure time for the two measurements.},
\begin{equation}
    \sigma\rs{N\rs{0}} = \sigma\rs{\Gamma} = \frac{1}{\sqrt{T}}.
\end{equation}
From Eq. \eqref{eq:errprop_2p} now we derive
\begin{equation}\label{eq:epszbcwithrho}
    \epszbc = N^2\rs{0}\biggl|\frac{8J\rs{1}(\pi \theta d\rs{2} / \lambda)J\rs{2}(\pi \theta d\rs{2} / \lambda)}{(\pi \theta d\rs{2} / \lambda)}\biggr|
    \frac{\Corr\Gamma\pm\sqrt{N^2\rs{0} + \Gamma^2(\Corr^2-1)}}{\Gamma^2-N^2\rs{0}}.
\end{equation}
As before, we can compare Eq. \eqref{eq:epszbcwithrho} with Eq. \eqref{eq:epspcoeff} and see that
\begin{equation}
\epszbc = \epsp N^2\rs{0}\frac{\Corr\Gamma\pm\sqrt{N^2\rs{0} + \Gamma^2(\Corr^2-1)}}{\Gamma^2-N^2\rs{0}}.
\end{equation}
In this case, the value of $\Corr$ can be calculated from the numerical values of $\epsp$ and $\epszbc$ computed from the simulations (the black dashed curve and the redder continuous curve in the lower panel of Fig. \ref{fig:etaeps_2p}). We find that
\begin{equation}\label{eq:corr_epszbc}
\Corr = \frac{\Gamma}{N\rs{0}} \frac{1}{\sqrt{1+(\Gamma/N\rs{0})^2}},
\end{equation}
which leads to a final expression for $\epszbc$,
\begin{equation}\label{eq:epszbc}
    \epszbc = \biggl|\frac{8J\rs{1}(\pi \theta d\rs{2} / \lambda)J\rs{2}(\pi \theta d\rs{2} / \lambda)}{(\pi \theta d\rs{2} / \lambda)}\biggr|
    \frac{1}{\sqrt{1+(\Gamma/N\rs{0})^2}}.
\end{equation}
It is worth nothing that $\etzbc$ and $\epszbc$ are independent of the value of the ZBC (we recall that $\Gamma = N\rs{0}\VisSq$). 

From Eq. \eqref{eq:corr_epszbc} we calculate that when the position of the second measurement tends to 0 ($\Gamma \to N\rs{0}$), $\Corr$ tends to $\sqrt{2}/2$ as in Eq. \eqref{eq:corr_etzbc}. This reflects the fact that when the position of the second measurement tends to the position of the first measurement, the two S/Ns tend to the same value and we return to the previous scenario. On the other hand, when the position of the second measurement tends to 1, $\Corr$ will tend to 0, breaking the correlation between the two parameters. As before, we can calculate the baseline of the second measurement needed to maximise the S/N: $d\rs{2}/d\rs{0} \simeq 0.550$ (slightly higher than in the case of known ZBC).

\subsection{Two measurements with different S/N or different integration time}

With the obtained results, we can now try to investigate the more general case of two different S/Ns or two different integration times. The previous results are very helpful to describe the general behaviour.

\begin{figure}
    \centering
    \includegraphics[width=1\columnwidth]{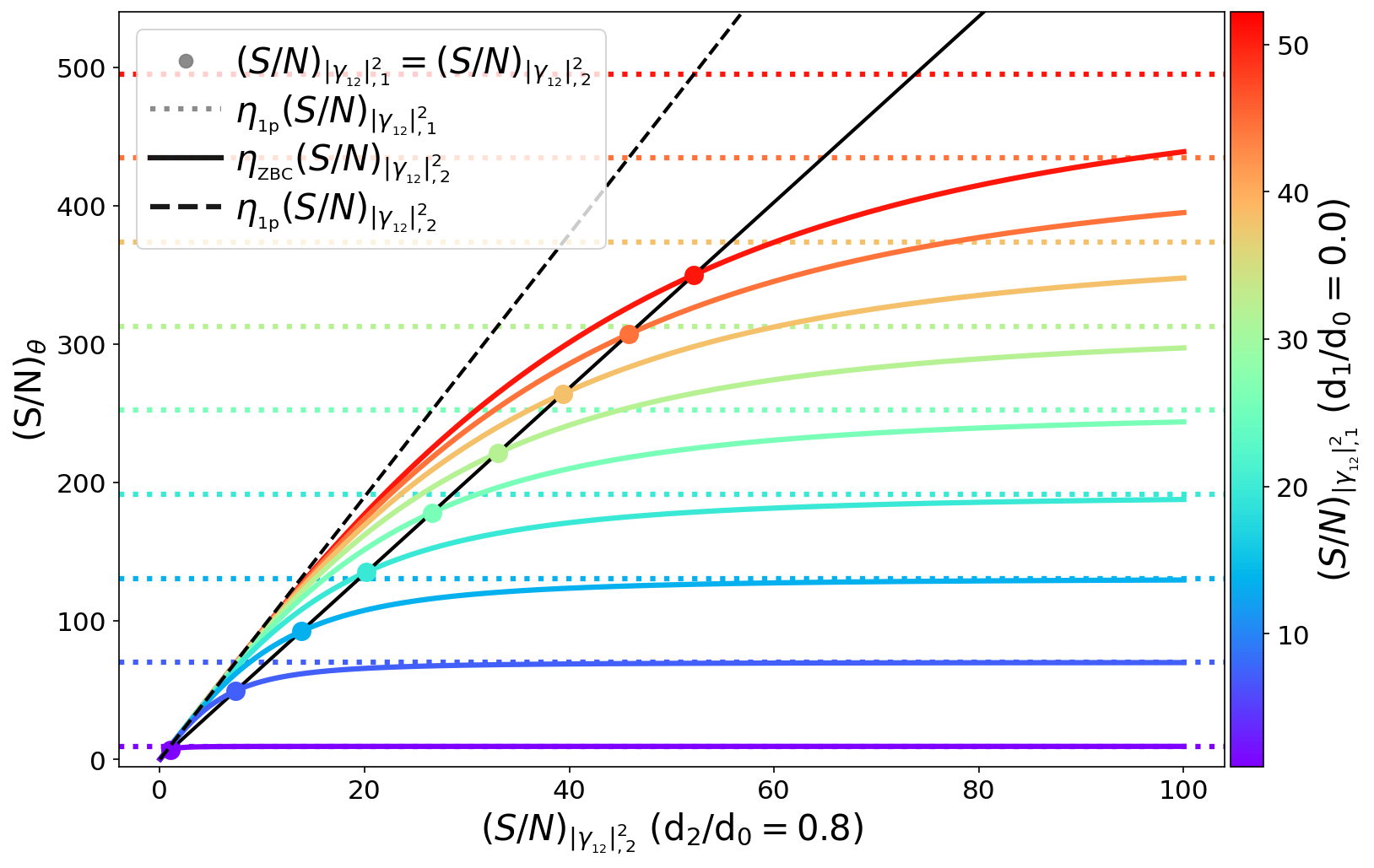}
    \caption{Trend of $\SNtheta$ as a function of $\SNgamsec$. The baselines of the first and second measurements are fixed to 0.0 and 0.8. The solid coloured curves correspond to different values of $\SNgamfir$. The black line shows the trend in case of equal S/N for the two measurements, the dashed black line the trend for the  single measurement scenario (Sect. \ref{sec:onepoint}), and the horizontal dotted lines represent the maximum limit of each of the corresponding coloured curves. The scenario in which we fix the second measurement is not shown because it is completely symmetric.}
    \label{fig:2p_diff_SN}
\end{figure}

Fig. \ref{fig:2p_diff_SN} shows $\SNtheta$ as a function of $\SNgamsec$ (the S/N of the longest baseline measurement). The different colour curves correspond to different values of $\SNgamfir$ (the S/N of the shorter baseline measurement). In this situation, we can make a direct comparison with the results obtained in the previous sections. 
The trend is no longer linear with $\SNgamsec$. On the other hand, the coloured measurements, corresponding to $\SNgamfir = \SNgamsec$, follow the linear trend resulting from Eq. \eqref{eq:etpp} (continuous black line).
When $\SNgamsec\ll\SNgamfir$, the curves asymptotically tend to the relation found for the scenario with the known ZBC, with $\SNtheta=\etp\SNgamsec$ (dashed black line). This means that for high values of $\SNgamfir$ and much lower values of $\SNgamsec$ , it is possible to approximate the trend of $\SNtheta$ with the relations that are valid for the known ZBC scenario.
When $\SNgamsec\gg\SNgamfir$, the curves asymptotically tend to the relation found for the scenario with the known ZBC, with $\SNtheta = \etp\SNgamfir$ (dotted coloured lines). This means that for high values of $\SNgamsec$ and much lower values of $\SNgamfir$ , it is possible to approximate the trend of $\SNtheta$ with the relations that are valid for the known ZBC scenario. However, in this case, the measurement known with the highest accuracy is that at larger baselines.
We thus observe that these results are symmetric. Swapping the S/N of the two measurements yields the same final $\SNtheta$, which essentially depends on the measurement with the lower S/N.
These results clearly show that the best strategy to optimise the S/N ratio of the observations is to have two measurements with similar S/N. It is not useful to increase one over the other indefinitely.

The same behaviour as $d\rs{1}/d\rs{0} = 0.0$ is found also for $d\rs{1}/d\rs{0} > 0.0$ (and for any value of $d\rs{2}/d\rs{0}$): an oblique and a horizontal asymptote are always present in plots like the one shown in Fig. \ref{fig:2p_diff_SN}, which means that the same prescriptions will still be valid.

\begin{figure}
    \centering
    \includegraphics[width=1\columnwidth]{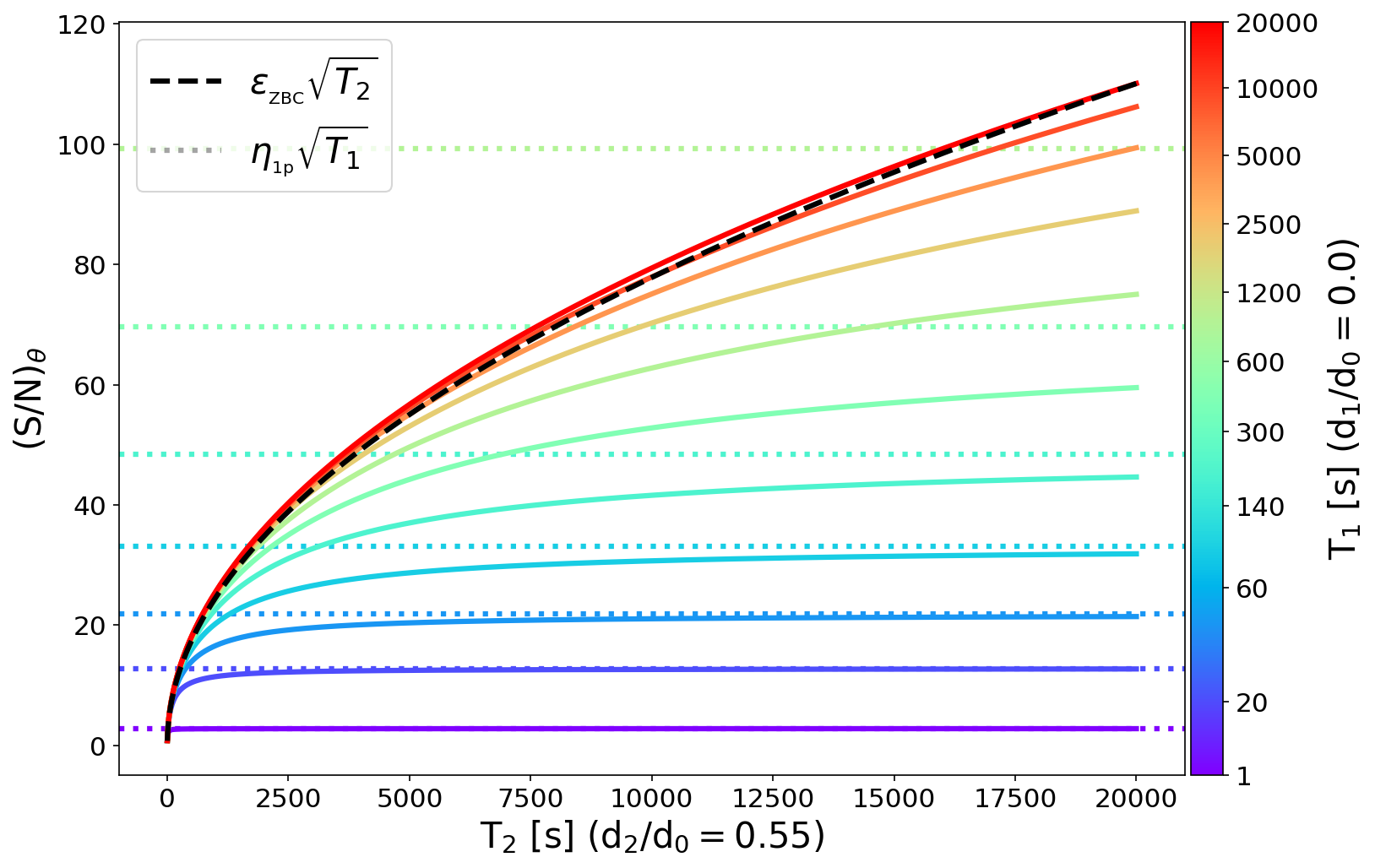}
    \includegraphics[width=1\columnwidth]{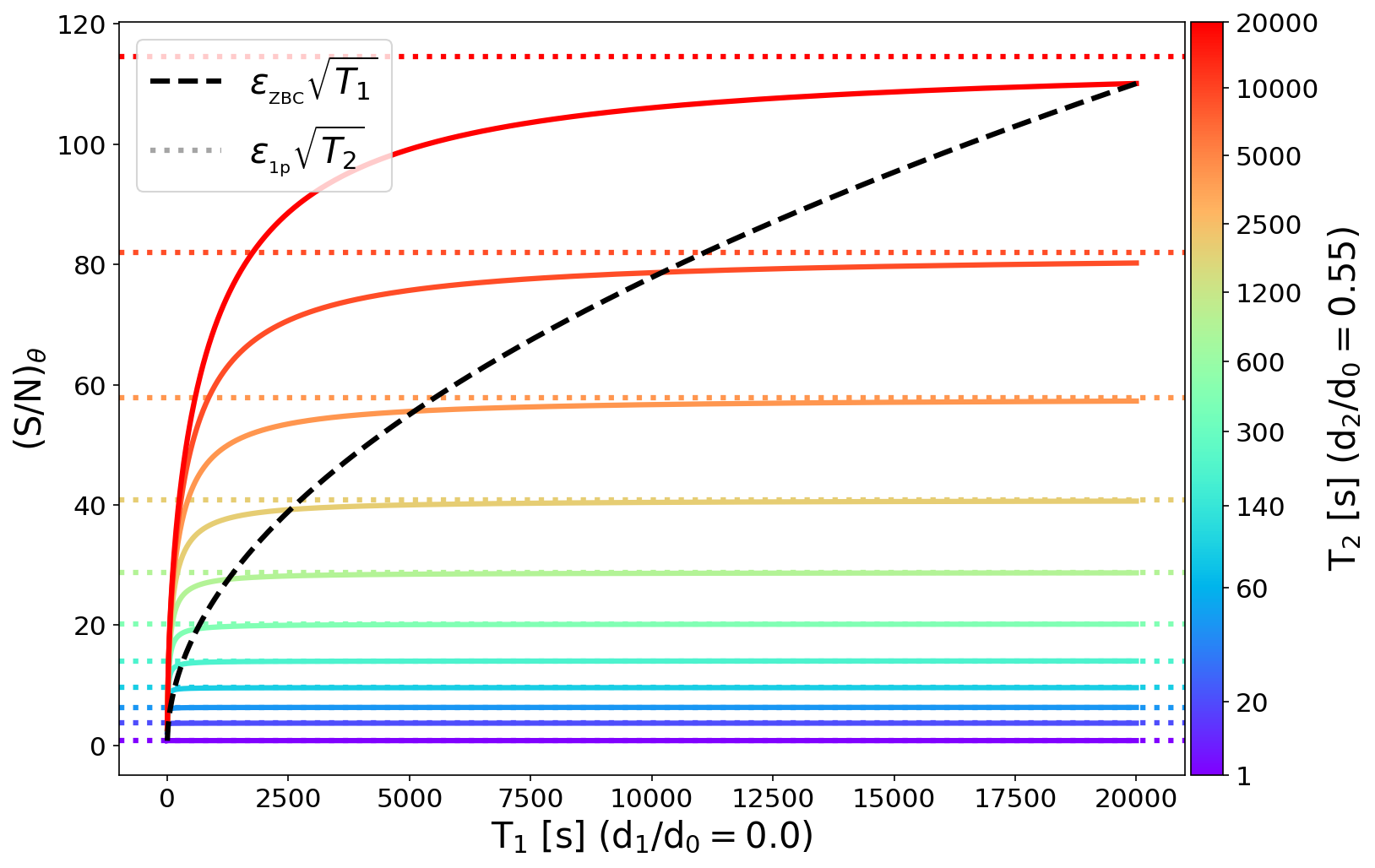}
    \caption{Trend of $\SNtheta$ as function of the integration time, with $T\rs{1}$ fixed (upper panel) and with $T\rs{2}$ fixed (lower panel). The dashed black curves show the trend in case of equal integration times, and the horizontal dotted lines show the maximum limit of each of the corresponding coloured curves.}
    \label{fig:2p_diffTs}
\end{figure}

We considered then the scenario with two different integration times, again showing as an example the results in the case where the normalised baseline of the first measurement is $d\rs{1}/d\rs{0} = 0.0$ and that of the second measurement is $d\rs{2}/d\rs{0} = 0.8$. In the top panel of Fig. \ref{fig:2p_diffTs} we report the results where the integration time $T\rs{1}$ of the first measurement is fixed (different coloured curves), while in the bottom panel we report the case where the integration time $T\rs{2}$ of the second measurement is fixed (differently coloured curves). Making a comparison with the previous results, we can say first that the different cases are no longer symmetric. The reason clearly is that different integration times are needed to obtain a similar S/N for the two measurements. 
Second, the general trend is no longer the square root of the integration time.
Fixing $T\rs{1}$, we find a situation analogous to the case shown in Fig. \ref{fig:2p_diff_SN}. Furthermore, for $T\rs{2}\leq T\rs{1}$ , the curves tend to follow the relation $\SNtheta \simeq \epszbc \sqrt{T\rs{2}}$ quite well, while for $T\rs{2}\geq T\rs{1}$ , the curves  grow slowly until they tend to a constant value that is defined by $\etp\sqrt{T\rs{1}}$. 
Fixing $T\rs{2}$ , we find that $\SNtheta$ quickly rises at the beginning and then suddenly changes slope around  a value close to $\epszbc\sqrt{T\rs{1}}$. After this time, $\SNtheta$ tends asymptotically to $\epsp\sqrt{T\rs{2}}$.
Thus, a higher $\SNtheta$ could be reached by increasing the exposure time $T\rs{2}$ of the second measurement, rather than increasing the exposure time $T\rs{1}$. This can be easily seen from the fact that in the lower plot, all the curves for $T\rs{1}<T\rs{2}$ have a higher $\SNtheta$ than in the relation for equal exposure times of the two measurements (dashed black curve).
We can therefore claim that when it is not possible to increase the exposure times of the two measurements, it is preferred to increase the exposure time of the measurement at larger baseline. This is opposite to what we showed before, where it was possible to improve the fit at the same level by increasing the S/N of one of the two measurements. 
However, it is worth nothing that there is no limit to the maximum $\SNtheta$  if it is possible to increase the two integration times.

\subsection{Three measurements}
\label{sec:threepoints}

When another measurement is added to the visibility curve, the final accuracy of the fit might be improve. The placement of this additional measurement is to be determined. In the following we show the results of two simple analyses we made considering different integration times.

\begin{figure}
    \centering
    \includegraphics[width=1\columnwidth]{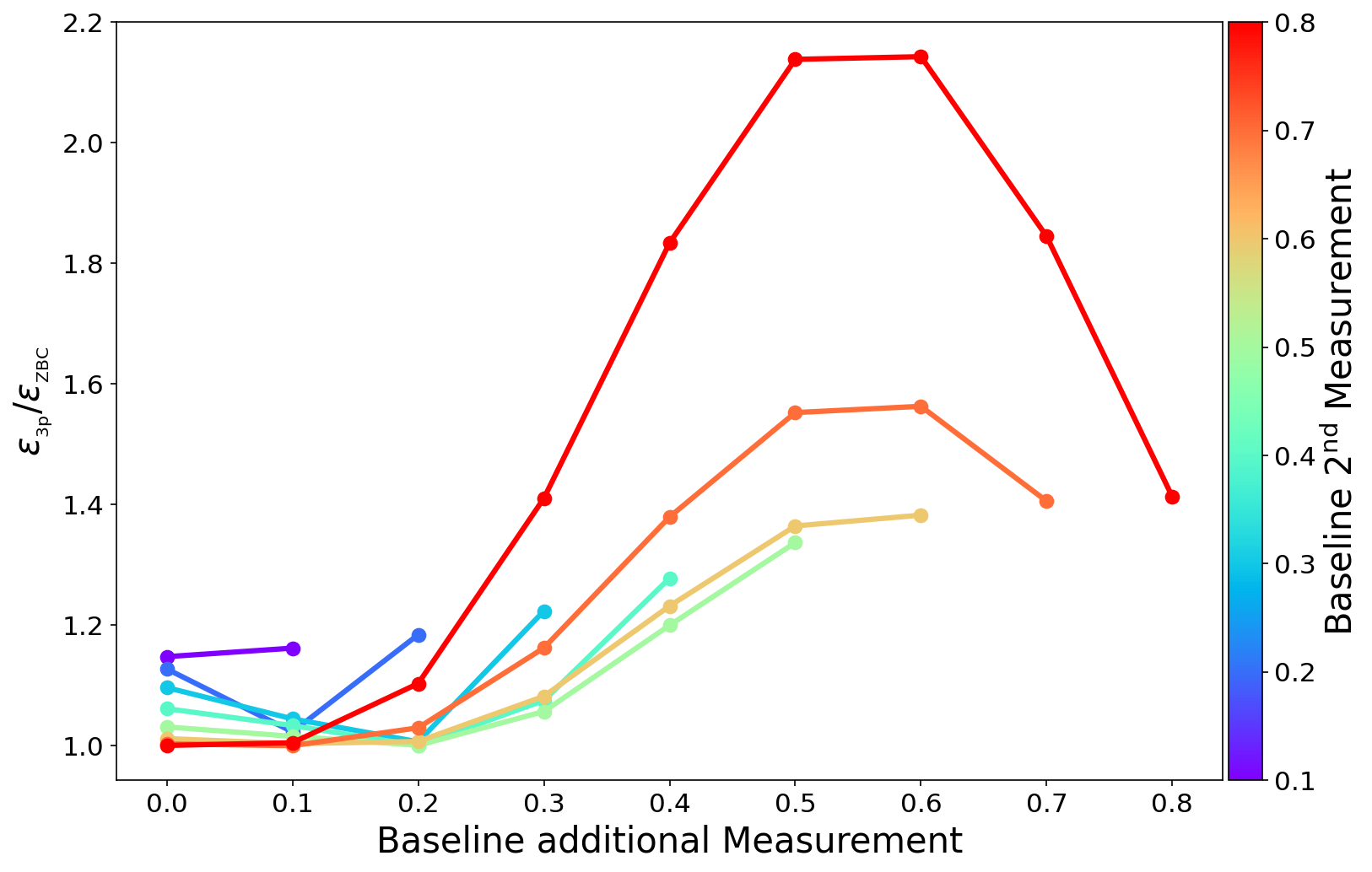}
    \caption{Ratio of $\epsppp$ and $\epszbc$ as a function of the baseline of the third measurement. The baseline of the first measurement is kept fixed to $d\rs{1}/d\rs{0} = 0.0$. The differently coloured curves correspond to different baselines of the second measurement. Results with a different baseline for the first measurement are similar to this case (with lower maximum values).}
    \label{fig:3p_coeffs}
\end{figure}

In the first analysis, we compared the results we obtained in the three measurements scenario with the two-measurement scenario, with the same integration times for all the measurements. As before, we refer to the special case in which we directly measure the ZBC value ($d\rs{1}/d\rs{0} = 0.0$). We assumed that the first measurement is at zero baseline, the second measurement at the largest baseline, and the third point at an intermediate baseline.
We find again that the $\SNtheta$ follows a square-root relation as a function of $T$. In analogy with the previous analyses, we can therefore write a relation of this type: $\SNtheta = \epsppp \sqrt{T}$. $\epsppp$ can be fitted directly to the curves and then be related to $\epszbc$ in the two-measurement scenario.
In Fig. \ref{fig:3p_coeffs} we show the ratio of the two coefficients as a function of the baseline of the third measurement. The ratio of the two coefficients basically shows the improvement of $\SNtheta$ compared to the two-measurement case.
As before, we find that the best results are achieved with at least one of the three measurements at a normalised baseline $d/d\rs{0} \sim 0.5-0.6$. While the point with the longest normalised baseline is at 0.8, having another measurement at $0.5-0.6$ will improve the results by more than a factor 2. In general, adding a measurement will always improve the final results because the statistics are increased.

\begin{figure}
    \centering
    \includegraphics[width=1\columnwidth]{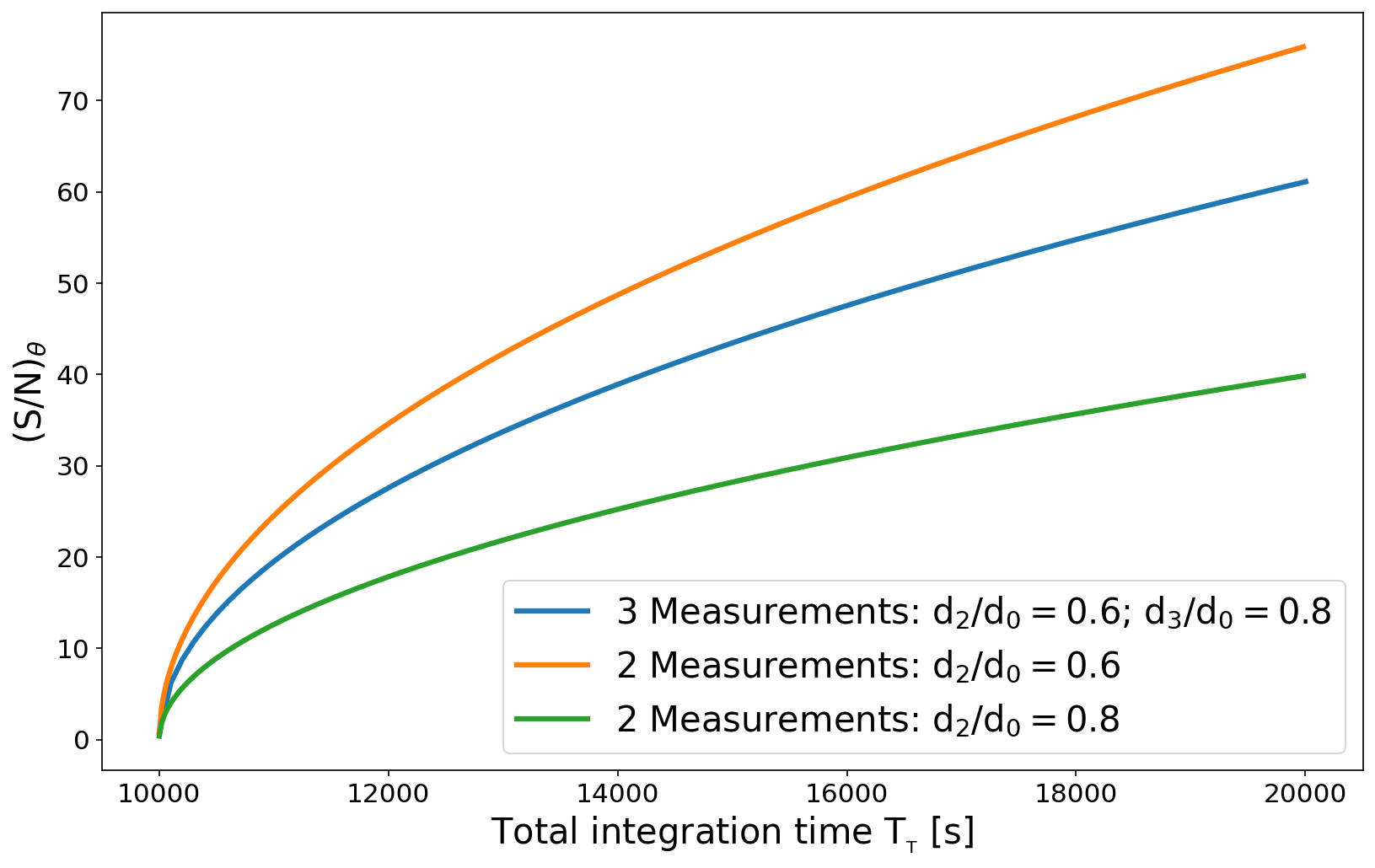}
    \caption{Trend of $\SNtheta$ as a function of the total integration time $T\rs{T}$ (the sum of the integration times of the measurements). The integration time for the first measurement is kept fixed to 10 ks. In this way, we compare results with the same total integration time (for the three-measurement scenario, the integration time  $T\rs{T}-T\rs{1}$ is split equally into the two remaining measurements). The blue curve correspond to the case in which we measure three measurements with baselines 0.0, 0.6, and 0.8. The orange curve correspond to the case in which we measure two points with baselines 0.0 and 0.6. The green curve correspond to the case in which we measure two points with baselines 0.0 and 0.8.}
    \label{fig:3pvs2p}
\end{figure}

On the basis of this result, the best strategy for optimising the available integration time might appear be to have one measurement at $0.5-0.6$ and one at 0.8. However, although in Fig. \ref{fig:3p_coeffs} we compare coefficients computed considering measurements with equal integration times, the total integration time $T\rs{T}$ (the sum of the integration times of all measurements) is different. In the three-measurement scenario, the total integration time is three and a half times that of the two-measurement scenario. 
Thus, to distinguish the best strategy, we carried out a second analysis in which we compared the results with the same total integration time. In Fig. \ref{fig:3pvs2p} we show an example of the trend of $\SNtheta$ as a function of the total integration time $T\rs{T}$. In this example $T\rs{T}$ is the same in all the cases. Again, we chose the case with $d\rs{1}/d\rs{0} = 0.0$, for which we set the integration time to a fairly high value ($T\rs{1}=10$ ks) to ensure that the final trend of $\SNtheta$ goes almost with the square root of the integration time of the other measurements (similarly to what we see with the red curve in the top panel of Fig. \ref{fig:2p_diffTs}).
In Fig. \ref{fig:3pvs2p} we show the curves of the three-measurement scenario with measurements at 0.6 and 0.8 (blue curve), of the two-measurement scenario with one measurement at 0.6 (orange curve), and of the two-measurement scenario with one measurement at 0.8 (green curve). 
 When the integration time for the first measurement is fixed, the integration time for the other measurement in the two-measurement scenarios is twice as long as the integration time for the two measurements in the three-measurement scenario. The maximum $\SNtheta$ is clearly reached when all the available integration time is spent on the measurement at 0.6 and is not divided between two measurements. The scenario in which all the time spent on the measurement at 0.8 is the worst case because we have no measurement at the best baseline.  We verified that in all cases it is better to spend all the integration time on a baseline ranging from 0.4 to 0.7 to obtain the best results.

\section{Realistic simulation}
\label{sec:simuluations}
All the analyses carried out in the previous section were made assuming $\alpha=1$ (see Eq. \ref{eq:SNgam}). We now wish to test the results considering a more realistic simulation. Our aim is to show that the previous equations can be used to predict the accuracy of the reconstructed angular size of a star. Following the simplified numerical procedure from \citealt{Dravins12} (Sect. 6.1), we produce a simulation of a hypothetical SII observation with two Cherenkov telescopes.

We considered a system that can observe a bright star at all possible baselines between 0 and $d\rs{0}$ in fully polarised light, measuring simultaneously the (unknown) ZBC value and another measurement at a longer baseline. The $\SNgam$  can now be written as
\begin{equation}\label{eq:finalSNgam}
    \SNgam = n \biggl(\frac{\lambda\rs{c}^2}{c\Delta\lambda}\biggr)\kappa \VisSq \sqrt{\frac{T}{dt}}, 
\end{equation}
where $n = \sqrt{\overline{N}\rs{A}\overline{N}\rs{B}}$ is the geometric mean of the average photon rates between the two detectors $A$ and $B$, $\lambda\rs{c}$ and $\Delta\lambda$ are the central wavelength and bandwidth of the optical filter, $\kappa$ is the overall efficiency of the system, $T$ is the integration time, and $dt$ is the sampling time.  
We then expect that $\SNtheta$ can be expressed as
\begin{equation}\label{eq:finalSNtheta}
    \SNtheta =  n \biggl(\frac{\lambda\rs{c}^2}{c\Delta\lambda}\biggr)\kappa \epszbc \sqrt{\frac{T}{dt}}, 
\end{equation}
where $\epszbc$ is given by eq. \eqref{eq:epszbc}. 
\begin{table}
\caption{Input parameters for the simulation shown in Fig. \ref{fig:simulations} and described in the text.}
    \label{tab:sim_setup}
    \centering
    \begin{tabular}{lc}\hline\hline
    \multicolumn{1}{l}{Parameter} & \multicolumn{1}{c}{Value} \\\hline
    Star size ($\theta$)     &  1 mas\\
    Photon rate geometric mean ($n$) & 20 Mct/s\\
    Central wavelength optical filter ($\lambda\rs{c}$) & 400 nm\\
    Optical filter bandwidth ($\Delta\lambda$) & 1 nm\\
    Sampling time ($dt$) & 1 ns\\
    System efficiency ($\kappa$) & 0.5\\
    Baseline first zero Bessel function ($d\rs{0}$) & $\sim100$ m\\
    Baseline first measurement ($d\rs{1}/d\rs{0}$) & 0.0\\
    Baseline second measurement ($d\rs{2}/d\rs{0}$) & 0.6\\
    Maximum observing time ($T$) & $\sim1.7$ h\\\hline
    \end{tabular}
\end{table}
All the parameters used in the simulation are reported in Table \ref{tab:sim_setup}. 

\begin{figure*}
    \centering
    \includegraphics[width=1\columnwidth]{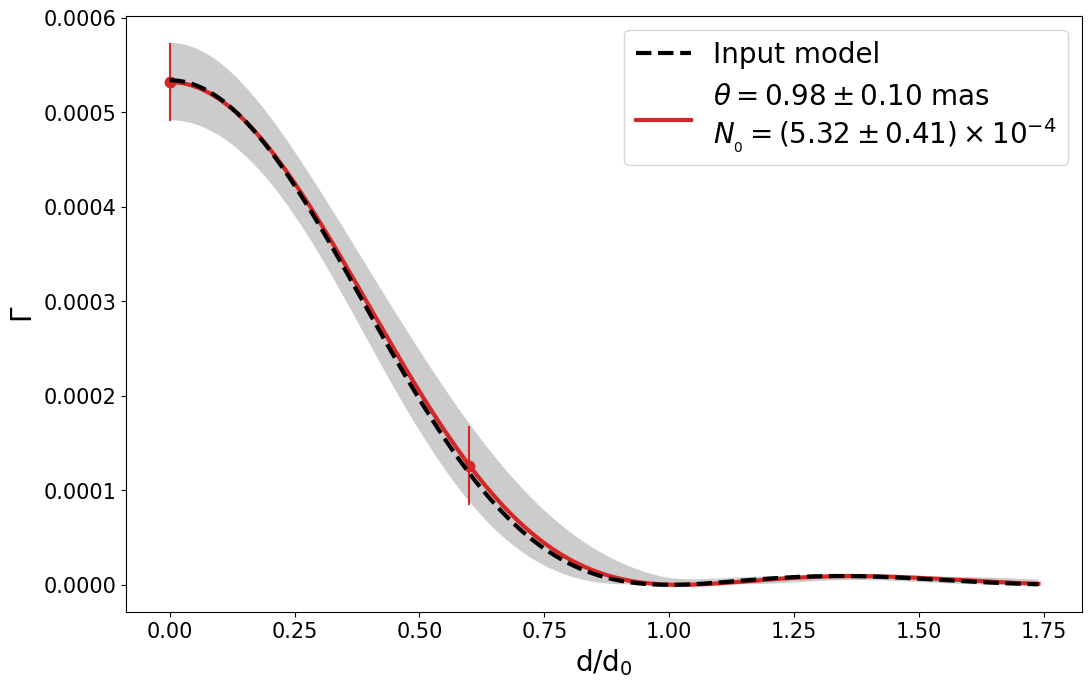}
    \includegraphics[width=1\columnwidth]{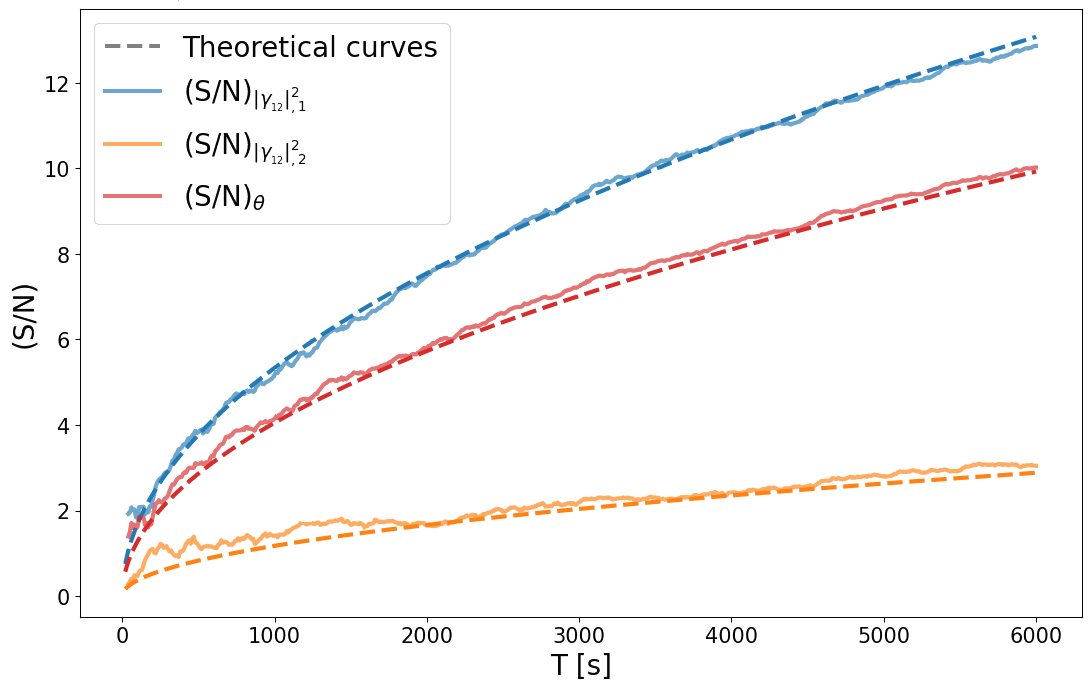}
    \caption{Results of the realistic simulation (averaged over ten realisations to reduce the statistical uncertainty of the single simulation). \textit{Left}: Fitted visibility function. The red measurements show the simulated data, the red curve shows the fitted model, the dashed black curve shows the input model (Table \ref{tab:sim_setup}) and the gray shaded area is the 1$\sigma$ confidence interval.  \textit{Right}: Trends of $\SNgam$ for the two measurements (blue and orange continuous curves) and of $\SNtheta$ (red continuous curve) as a function of the integration time. The measured curves are compared to the theoretical predictions (dashed curves) from Eq. \eqref{eq:finalSNgam} and \eqref{eq:finalSNtheta}.}
    \label{fig:simulations}
\end{figure*}

Figure \ref{fig:simulations} shows the results of this simulation (averaged over ten realisations to reduce the statistical uncertainty of the single simulation). In the left panel, we plot the final fitted visibility function (red curve) together with the input model (dashed black curve) and two simulated measurements (red points). The reconstructed angular size agrees with the input size, and $\SNtheta$ is equal to $\sim9.99$. The theoretical value computed from Eq. \eqref{eq:finalSNtheta} is $\sim9.93$, which is very close to the measured value. In the right panel, we show the measured trend of $\SNgam$ for the two simulated measurements (blue and orange curves) and the trend of $\SNtheta$ (red curve) as a function of the integration time. The dashed curves are the corresponding trends computed from Eq. \eqref{eq:finalSNgam} and Eq. \eqref{eq:finalSNtheta}. This simulation clearly shows that the theoretical values calculated through the expressions derived in this study and the actual measured value of the error on the angular size of a star inferred from the simulations agree remarkably well. 

\section{Discussion and conclusions}
\label{sec:conclusion}

We studied the accuracy of the reconstructed angular size of a star that can be achieved through SII observations. This study was carried out to understand how the SII data acquisition can be optimised. Because Cherenkov telescope arrays have fixed positions, changing the baseline of the observation means that the stars are observed at different positions in the sky (and therefore at different times during one night or during different nights in the year). When a guess of the angular size of the observed star is known in advance (e.g. through previous observations or from some stellar model predictions), some practical prescription of how to optimise the observing plans will be very useful.

We recall that we introduced some simplifications. We adopted an approximate model for the stellar brightness profile, a uniform disk model (Eq. \ref{eq:uniformdisk}), and  a least-squares algorithm for the data fitting. 
However, for the purpose of deriving a first-order estimate of the error accuracy, this simplified approach provides very useful information. The results that can be obtained with more sophisticated methods are not expected to be much more accurate than our estimations. A valid objection, however, might be that in reality, the uniform disk model does not fit the true stellar size because of limb-darkening effects and the presence of hot or cold spots on the stellar surface. This is another reason why we focused only on the fit of the first peak of the visibility function because the effects of a deviation from a uniform disk become strong from the second peak.

We found that knowing the ZBC value, as in the case of systems that perform SII in photon-counting mode, allowed measuring a stellar size using just one measurement on the visibility curve. When the value of $\SNgam$ for this single measurement was considered alone, the best results were obtained when it was taken at the longest possible baseline. On the other hand, when we considered the integration time needed to achieve a certain accuracy, we found that it is better to have a measurement around $d/d\rs{0} \sim 0.527$ because at a larger baseline, the time needed to obtain an acceptable $\SNgam$ increases faster than the improvement in the precision of the fit.
Similarly to the case where the ZBC value is known, it is then better to obtain a second measurement at $d/d\rs{0} \sim 0.550$ to optimise the results. This position tends to move towards larger baselines as the baseline of the first measurement increases. 
This is somehow similar to what has also been found by \cite{Rai22}, who demonstrated in the more complex situation of a close binary system that the best results are obtained by taking measurements at certain specific positions on the visibility curve.

We have found that similar $\SNgam$ or similar integration times for different measurements of the visibility curve will basically be the best observation strategy. Spending too much time on one of the measurements tends to saturate the maximum achievable $\SNtheta$ (horizontal limits in Fig. \ref{fig:2p_diff_SN} and \ref{fig:2p_diffTs}). If the system in use does not allow observing two (or more) measurements at the same time, and if only one of the two measurements can be observed for a long time, it is then better to spend more time at a longer baseline (because more time is required to obtain a good $\SNgam$). However, it is advisable to try to have similar integration times for the measurements because $\SNtheta$ can improve more.
When we tried to determine where another measurement should best be added, we found that it is better to add it at a normalised baseline around $0.5-0.6$, regardless of the position of the two other measurements. The best strategy is again measuring the ZBC value and two, three, or $n$ measurements around a baseline of $0.5-0.6$ (see Figs. \ref{fig:3p_coeffs} and \ref{fig:3pvs2p}).

The most useful results of this work certainly include the analytical expressions describing the trend of $\SNtheta$ as a function of $\SNgam$ or the integration time. The expressions for $\etp$, $\epsp$, $\etzbc$, and $\epszbc$ can be directly used to obtain estimates of $\SNtheta$, similarly to what is done usually for $\SNgam$. We tested the robustness of these results with a detailed simulation in which we adopted the expression for $\SNgam$ for photon-counting SII. The behaviour of $\SNtheta$ perfectly agreed with our predictions based on Eq. \eqref{eq:finalSNtheta}.

To summarise, it is not always necessary to measure the visibility curve at many different positions to obtain the best results. Instead, the best possible results can be obtained by taking a series of measurements at the ZBC and at a baseline half-way between zero and the position of the first zero of the visibility curve. This consideration can have a significant impact for planning the observations with the current class of Cherenkov SII instruments, as well as for the forthcoming instruments. If a system can measure the ZBC value (or if it is known in advance), it will then be possible to plan the observations by predicting the achievable accuracy without having to carry out precise simulations.

\section*{Acknowledgements}

We would like to thank Juan Cortina and Prasenjit Saha for their reading of the draft of this manuscript and for all the useful interactions we had.
This research has been funded by the University of Padova under the project "BIRD NALE\_SID19\_01".
This research made use of the following PYTHON packages: MATPLOTLIB \citep{Matplotlib2007}, NUMPY \citep{Numpy2011}, SCIPY \citep{Scipy2020}, PANDAS \citep{Pandas2010}.

\bibliographystyle{aa}
\bibliography{SII}
%
%

\end{document}